\documentclass[draft]{agujournal2019}
\usepackage{url} 
\usepackage{lineno}
\usepackage[inline]{trackchanges} 
\usepackage{soul}
\usepackage{xcolor} 
\usepackage{pdflscape}
\usepackage{amssymb}
\usepackage{amsmath}
\draftfalse

\journalname{Journal of Advances in Modeling Earth Systems (JAMES)}

\begin{document}

%
%

\title{Non-Linear Dimensionality Reduction with a Variational Encoder Decoder to Understand \\   
Convective Processes in Climate Models}

%
%




\authors{Gunnar Behrens \affil{1,2}, Tom Beucler \affil{3}, Pierre Gentine \affil{2,4}, Fernando Iglesias-Suarez \affil{1}, Michael Pritchard \affil{5}, Veronika Eyring \affil{1,6}}


\affiliation{1}{Deutsches Zentrum für Luft- und Raumfahrt (DLR), Institut für Physik der Atmosphäre, Oberpfaffenhofen, Germany}
\affiliation{2}{Department of Earth and Environmental Engineering, Columbia University, New York, NY 10027, USA}
\affiliation{3}{Institute of Earth Surface Dynamics, University of Lausanne, Lausanne, VD 1015, Switzerland}
\affiliation{4}{Earth Institute and Data Science Institute, Columbia University, New York, NY 10027, USA}
\affiliation{5}{Department of Earth System Science, University of California Irvine, Irvine, CA, USA}
\affiliation{6}{University of Bremen, Institute of Environmental Physics (IUP), Bremen, Germany}



\correspondingauthor{Gunnar Behrens}{Gunnar.Behrens@dlr.de}




\begin{keypoints}
\item A variational encoder-decoder (VED) can predict sub-grid-scale thermodynamics from the coarse-scale climate state.
\item The VED's latent space can distinguish convective regimes, including shallow/deep/no convection.
\item The VED's latent space reveals the main sources of convective predictability at different latitudes.
\end{keypoints}

%
%

%
%


\begin{abstract}

Deep learning can accurately represent sub-grid-scale convective processes in climate models, learning from high resolution simulations. However, deep learning methods usually lack interpretability due to large internal dimensionality, resulting in reduced trustworthiness in these methods. Here, we use Variational Encoder Decoder structures (VED), a non-linear dimensionality reduction technique, to learn and understand convective processes in an aquaplanet superparameterized climate model simulation, where deep convective processes are simulated explicitly. We show that similar to previous deep learning studies based on feed-forward neural nets, the VED is capable of learning and accurately reproducing convective processes. In contrast to past work, we show this can be achieved by compressing the original information into only five latent nodes. As a result, the VED can be used to understand convective processes and delineate modes of convection through the exploration of its latent dimensions. A close investigation of the latent space enables the identification of different convective regimes: a) stable conditions are clearly distinguished from deep convection with low outgoing longwave radiation and strong precipitation; b) high optically thin cirrus-like clouds are separated from low optically thick cumulus clouds; and c) shallow convective processes are associated with large-scale moisture content and surface diabatic heating. Our results demonstrate that VEDs can accurately represent convective processes in climate models, while enabling interpretability and better understanding of sub-grid-scale physical processes, paving the way to increasingly interpretable machine learning parameterizations with promising generative properties.

\end{abstract}

\section*{Plain Language Summary}
Deep neural nets are hard to interpret due to their hundred thousand or million trainable parameters without further postprocessing. We demonstrate in this paper the usefulness of a network type that is designed to drastically reduce this high dimensional information in a lower-dimensional space to enhance the interpretability of predictions compared to regular deep neural nets. Our approach is, on the one hand, able to reproduce small-scale cloud related processes in the atmosphere learned from a physical model that simulates these processes skillfully. On the other hand, our network allows us to identify key features of different cloud types in the lower-dimensional space. Additionally, the lower-order manifold separates tropical samples from polar ones with a remarkable skill. Overall our approach has the potential to boost our understanding of various complex processes in Earth System science.

%
%

\section{Introduction}
%


Earth System Models (ESM) are essential tools to investigate projected changes in precipitation patterns due to different climate scenarios. However, the atmospheric component of traditional ESMs, an atmospheric Global Circulation Model (GCM) with resolution of around 100 km, cannot directly simulate precipitation-generating convective processes, as they occur on scales of a few kilometers, so on much smaller length scales than the grid resolution \cite{Randall2003, Bony2015}. Therefore, GCMs rely on parameterizations to represent the effect of convective sub-grid-scale processes on the large-scale resolved state \cite{Randall2003, Randall2013}. However, the models exhibit large persisting systematic biases such as the presence of a Double Inter Tropical Convergence Zone (ITCZ, a zonal band of strong precipitation in the tropics that forms the ascending branch of the Hadley cell), uncertainties in shortwave cloud radiative forcing, or an over- or underestimation of cloud cover in certain regions as can be seen in recent phases of the Coupled Model Intercomparison Project \cite<CMIP,>{Bock2020}.  

A suitable and well-established alternative is the use of Storm Resolving Models (SRM, grid size $\sim$ 4 km), where a large fraction of convective processes (e.g. deep convection) is directly simulated. SRMs alleviate a number of issues in GCMs, such as the diurnal cycle, the representation of convective aggregation, or the variability and intensity of precipitation \cite{Stevens2020}. Nevertheless, SRM simulations still rely on parameterizations of fine-scale processes, which strongly affect precipitation formation (microphysics) and the onset of convective processes (turbulence). Due to the high computational costs of SRM runs, model iterations are limited to a sequence of months up to a few years, and long-term climate projections will remain unfeasible within the next decade on such fine resolutions.

Recent advances in machine learning (ML) have shown great potential in learning sub-grid-scale convective processes and are a promising approach to improve parameterizations in GCMs. Deep feed-forward neural nets showed great accuracy learning convective processes explicitly represented in a superparameterized aquaplanet simulation \cite{Gentine2018}, and successfully replaced the physics package of the GCM leading to stable prognostic ML-based simulations \cite{Rasp2018}. The ML-based model showed substantial improvements simulating both the mean climate and its variability, as represented by the superparameterized model, compared to the host GCM with conventional parameterizations. Similar advances in the prognostic skill, where the ML approach is coupled to the dynamical core of a circulation model, of global precipitation distributions on an aquaplanet were achieved by \citeA{Yuval2020}, with random forests and neural networks. Beyond aquaplanets, \citeA{Mooers2021} showed that feed-forward neural nets also skilfully reproduce the superparameterization (SP) in the presence of real topography based on offline tests, where the ML approach is evaluated against test data, without implementing the resulting representation back into the GCM. \citeA{Han2020} likewise demonstrated the potential to learn SP with residual neural nets based on real topography data. \citeA{Wang2021} showed the possibility to achieve a stable decade-long hybrid ML-Community Atmosphere Model run with real topography based on an ensemble of multiple residual neural nets that separately emulate convective heating, moistening, downwelling solar radiation and radiative fluxes affected by convection, albeit with some distortions of time-mean tropical rainfall bands. 

Most of these studies used neural net architectures with several hidden layers and hundreds of thousands or millions of degrees of freedom (weights and biases of the networks), with the exception of \citeA{Yuval2020}. Therefore, for all reviewed cases, quantifying the influence of one large-scale climate variable (input) on an emulated sub-grid-scale variable (output) remains challenging without well-suited state-of-the-art interpretability or attribution methods, especially for such high-dimensional regression tasks \cite{Mamalakis2021}. This is due to these machine learning algorithms' large internal variability, and  clearly limits the trustworthiness of the reproduced sub-grid-scale variables and the reproduced variability. 

In this context, it is natural to wonder whether the use of lower-order models with a smaller latent manifold might prove a promising strategy to overcome the reliance on computationally expensive attribution methods. Our goal is to simplify the interpretation of reproduced convective processes and to provide physical interpretation of the learnt relationships, and more generally to reduce the effective dimensionality and build trust in the estimated emulation of convective processes. The purpose of this study is thus to explore whether Variational (Auto)Encoder (VAEs) Decoder structures \cite{Kingma2014} can realistically reproduce convective processes, while enhancing the interpretability of the complex interaction between convection and driving large-scale conditions.

VAEs have only begun to be explored in the atmospheric sciences. Initially \citeA{Alberdi2018}, showed the potential of VAEs to compress non-linear and chaotic data of the Lorenz 96' model \cite{Lorenz1996}, which is an ansatz for the turbulent convective nature of the atmosphere. VAEs proved to be powerful tools for the identification of different phases of Northern Hemispheric polar vortex in reanalysis data \cite{Krinitskiy2019}. They demonstrated the applicability of VAEs for common spatio-temporal climate data sets and their advantage compared to more standard linear approaches like Empirical Orthogonal Functions. A further step towards the use of VAEs has been the objective self-supervised classification of convective regimes based on the fine (kilometer-scale) details of explicitly resolved updrafts in global simulations \cite{Mooers2020}. Their VAE identified different tropical convective regimes based on embedded cloud-scale vertical velocity profiles within the embedded subdomains of a global SP simulation. They further showed that their VAE is a powerful approach to detect anomalies like tropical convective extremes and geographically rare forms of dry, continental convection in climate data sets with strong spatio-temporal variability \cite{Mooers2020}.

Here, we use a variational network to investigate the effective dimensionality of the convective parameterization problem, as well as interpret its latent space to delineate convective regimes and large-scale drivers of convective processes. Previous studies without ML approaches explored the interaction of convection and the large-scale climate conditions \cite{Derbyshire2004} or convective regimes \cite{Frenkel2012,Frenkel2013,Frenkel2015,Huaman2017} mostly in the tropics and subtropics. While the art and science of interpreting latent spaces is in its infancy, we will demonstrate that one promising method is to leverage the generative modeling capabilities by direct manipulation of the latent manifold. For our variational network, this reveals different convective regimes and how they are connected to driving large-scale conditions (temperature, specific humidity and radiative processes). For instance, we explore whether the geographic region of a GCM sample can be inferred solely based on its latent space position.  

The paper is organized as follows. Section 2 describes the climate simulation and machine learning approach used in this study. Section 3 focuses in its first part on the deterministic skill of a variational network for sub-grid-scale SP variables (decoding capabilities) and in the second part on the physical interpretability and meaningfulness of the resulting latent space (encoding capabilities of large-scale climate conditions and convective processes). Section 4 leverages the variational network's latent space to explore the drivers of different convective regimes. Finally, Section 5 provides a discussion and summary of the enhanced interpretability of convective processes via variational networks, as well as an outlook of such generative ML approaches in the context of new hybrid climate models.

\section{Data and Methods}

\subsection{Data: Superparameterized Aquaplanet Simulation}

We use a 2-year aquaplanet simulation of the superparameterized Community Atmosphere Model v3.0 (SPCAM) \cite{Collins2006,Khairoutdinov2005} under the configuration of \citeA{Pritchard2014} in which Sea Surface Temperatures (SST) were imposed following a realistic zonally symmetric distribution \cite{Andersen2012}. The SST maximum in the tropics is slightly displaced to 5$^{\circ}$ N and decreases meridionally towards the poles to reduce exact equatorial symmetry. The solar forcing is fixed to Austral Summer conditions (no seasonal variability), but includes diurnal variability.
The model has a coarse horizontal resolution corresponding to a typical grid size of 300 km near the equator. The vertical axis extends from the surface to $\sim$ 40 km (3.5 hPa) following a hybrid coordinate with 30 levels (22 levels below 100 hPa). The GCM uses a 30-minute time step. Following \citeA{Pritchard2014b}, the superparameterized (SP) component consists of 8 nested 2D columns oriented meridionally on the same vertical axis and with a sub-grid size of 4 km \cite{Grabowski2001,Khairoutdinov2001}. Deep convection is explicitly resolved every 20 seconds and a Smagorinsky 1.5-order turbulence closure, and a one-moment microphysics parameterization \cite{Khairoutdinov2003} are used. SPCAM in this configuration yields a realistic reproduction of the ITCZ and tropical wave-spectra with a pronounced Madden-Julian-Oscillation (MJO)-like signal, as well as improved precipitation distributions compared to the host GCM \cite<CAM,>{Pritchard2014b}. However, this SPCAM setup neglects momentum transport, and for our approach, we sidestep the SP of cloud ice and water sources and sinks and instead emulate their radiative consequences through the total diabatic heating, as in \citeA{Rasp2018}.

\subsection{Model: Variational Encoder Decoder}

\begin{figure}[ht]
    \centering
    \includegraphics[width=13.9cm]{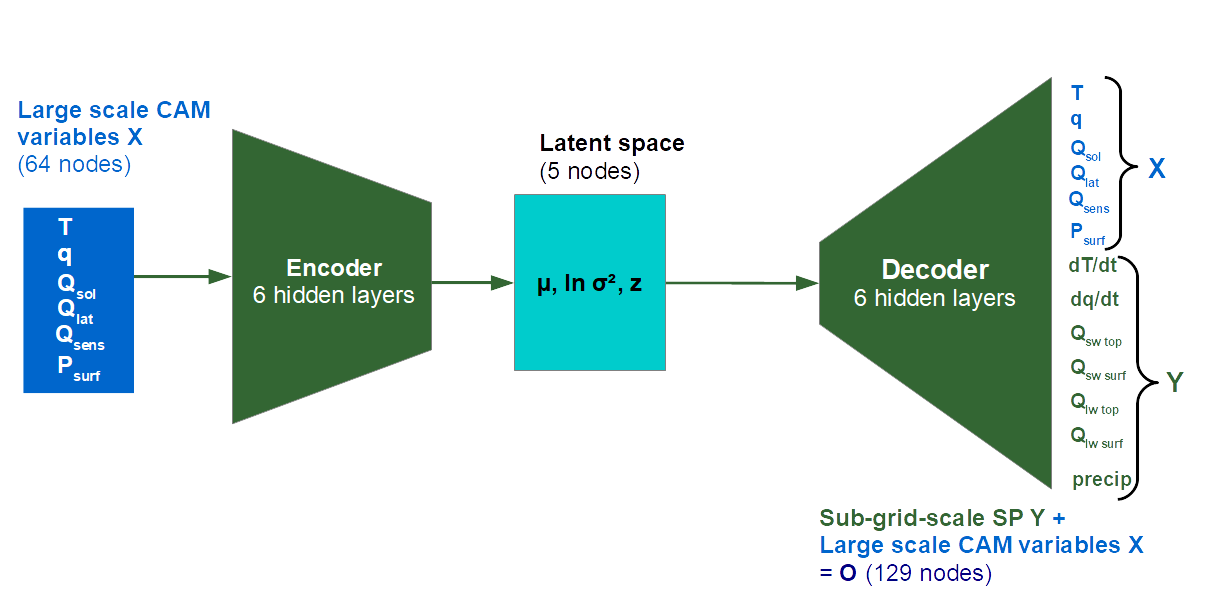}
    \caption{Schematic of the constructed VED which uses large scale CAM variables to investigate simulated sub-grid-scale convective processes of SP. The latent space consists of mean $\mathbf{\mu}$, a logarithmic variance $\mathbf{\ln \sigma^2}$ layer and the latent variables \textbf{z}. The output data \textbf{O} of the decoder includes a reconstruction of the input data \textbf{X} to the encoder to encourage a latent space that can additionally compress the large- scale climate variables, in addition to their mapping to the target sub-grid-scale fields \textbf{Y}. }
    \label{fig:VAE_schematic}
\end{figure}

We develop a variational encoder decoder (see schematic in Figure \ref{fig:VAE_schematic}) to holistically learn sub-grid-scale processes in SPCAM. VAEs traditionally reproduce their inputs, e.g., learning a mapping from large-scale variables to themselves. Here, our goal is to map large-scale to sub-grid-scale variables. Therefore, we adopt a variational encoder decoder (VED) architecture to include the emulation of sub-grid-scale variables. We include convection, turbulence, and radiation by simultaneously predicting the total diabatic heating and moistening tendencies alongside a decoded reconstruction of the relevant input data that summarize local large-scale state information prior to radiative-convective adjustment. Compared to deep feed-forward neural nets, the variational encoder decoder enhances the interpretability of convective processes and how they are connected to the driving large-scale climate via its latent space of reduced dimensionality. Regarding the input fields (\textbf{X}), we closely mirror the established precedent of \citeA{Rasp2018} by using profiles of specific humidity \textbf{q(p)} in $\frac {kg}{kg}$ and temperature \textbf{T(p)} in K on 30 vertical levels each, as extracted from the end of the host model dynamics or the beginning of the physics package. \textbf{X} additionally includes the scalar values of solar insolation \textbf{Q$_{sol}$} in $\frac{W}{m^2}$, surface latent heat flux \textbf{Q$_{lat}$} in $\frac{W}{m^2}$ and surface sensible heat flux \textbf{Q$_{sens}$} in $\frac{W}{m^2}$, and surface pressure \textbf{P$_{surf}$} in Pa. That is, \textbf{X} is a
concatenation of these two vectors and four scalars, [\textbf{q(p)}, \textbf{T(p)}, \textbf{Q$_{sol}$}, \textbf{Q$_{lat}$}, \textbf{Q$_{sens}$}, \textbf{P$_{surf}$}], into a 64-element input vector. The variational encoder decoder is trained to predict \textbf{O}, which combines the reconstruction of the same large-scale input data (as described above) with the sub-grid-scale process rate output fields targeted by \citeA{Rasp2018} \textbf{Y} (i.e., a parameterization): vertical profiles of total diabatic specific humidity tendency \textbf{dq(p)/dt} in $\frac {kg}{kg \times s}$ and total diabatic temperature tendency \textbf{dT(p)/dt} in $\frac{K}{s}$ defined on 30 pressure levels, as well as  scalar values for shortwave and longwave radiative heat fluxes at the model top (\textbf{Q$_{sw \ top}$} and \textbf{Q$_{lw \ top}$}) and at the surface (\textbf{Q$_{sw \ surf}$} and \textbf{Q$_{lw \ surf}$}) in $\frac{W}{m^2}$, and precipitation rate \textbf{precip} in $\frac{m}{s}$. 
The full predicted vector
\textbf{O} = [\textbf{dq(p)/dt}, \textbf{dT(p)/dt}, \textbf{Q$_{sw \ top}$}, \textbf{Q$_{sw \ surf}$}, \textbf{Q$_{lw \ top}$}, \textbf{Q$_{lw \ surf}$}, \textbf{precip}, \textbf{q(z)}, \textbf{T(z)}, \textbf{Q$_{sol}$}, \textbf{Q$_{lat}$}, \textbf{Q$_{sens}$}, \textbf{P$_{surf}$}] has a dimension of 129. 

As it will be the main ML model used in this study, we henceforth abbreviate the variational encoder decoder structure simultaneously predicting sub-grid-scale convective processes and large-scale climate conditions to ``VED'' for simplicity. A prior experiment with a VED$_{X \rightarrow Y}$ that was trained on \textbf{X} to predict \textbf{Y}, similar to the established precedent of \citeA{Rasp2018}, does not encode the large-scale climate variables \textbf{X} as much in its latent space compared to VED. This limited our ability to gain insight into convective predictability with VED$_{X \rightarrow Y}$ (see supporting material S.3A and Figure S16 for details). In contrast the combined reproduction of sub-grid-scale processes and large-scale climate variables with VED together with our generative modeling method allows us to explore convective regimes and corresponding large-scale climate conditions.

The encoding part of the VED (Encoder) consists of 6 hidden layers, which progressively reduce the dimensionality from 463 nodes in the first hidden layer down to 5 nodes (the latent variables) in the latent space. These values were chosen following a formal hyperparameter search (see supporting material S.1). We will test the sensitivity of emulations of the VED with respect to the number of latent nodes in section 3 in detail. In the following we will refer to one distinct latent variable in the context of the network architecture as ``latent node''. While we will use the notation ``latent space'' for the manifold spanned by all latent variables. Within this latent space, the mean $\mathbf{\mu}$ and logarithmic variance $\mathbf{\ln \sigma^2}$ are computed for each node, where $\mathbf{\sigma}$ is the standard deviation of the posterior \cite{Kingma2014}. Then a so-called `reparameterization trick' \cite{Kingma2014} is utilized to map the original distribution based on $\mathbf{\mu}$ and $\mathbf{\ln \sigma^2}$ onto an isotropic gaussian distribution. We used the $\mathbf{\ln \sigma^2}$ instead of $\mathbf{\sigma^2}$ for the construction of the network to simplify the reparameterization and the computation of the VED loss.
The resulting latent variables \textbf{z} (5 dimensions) are used to investigate convective processes and drivers of convective predictability. Henceforth we will use the notation ``latent dimension'' to describe the subspace spanned by one particular latent variable. We will show in section 4 that characteristic convective regimes and large-scale climate states are encoded in \textbf{z}. The latent variables \textbf{z} are the only input fed to the decoding part of the VED (Decoder), which reconstructs both large-scale and sub-grid-scale fields. In the decoder, the dimensionality is progressively increased to 463 in the last hidden layer before the 129-node output layer. We use the rectified linear unit (relu) as activation function of all hidden layers of the Encoder and Decoder except for the Decoder output layer, where we use an exponential linear unit (elu) based on prior hyperparameter testing (see S.1). In the latent space, $\mathbf{\mu}$ and $\mathbf{\ln \sigma^2}$ are linearly activated, whereas for the latent variables \textbf{z} we call the reparameterization function. In summary, the Encoder and Decoder of the VED consist of 388,440 and 418,469 total trainable parameters, respectively. 

We train the VED over 40 epochs (number of iterations through training data), during which the weights and biases are updated to minimize the VED loss function (see Equation 1). 

\begin{linenomath*}
 \begin{equation}
 \mathrm{VED} \ \mathrm{loss}  =  \mathrm{reconstruction} \  \mathrm{loss} + {\lambda} \ \mathrm{KL} \ \mathrm{loss}
 \end{equation}
 \end{linenomath*}

The loss function is the sum of a reconstruction and a Kullback-Leibler (KL, Equation 3) loss term. The first term measures the mean-square error (MSE, Equation 2) between the predicted (\textbf{O$^{emul}$}) and the ground truth data (\textbf{O}).

\begin{linenomath*}
 \begin{equation}
 \mathrm{reconstruction} \  \mathrm{loss} = {{ \frac{1} {M}} \times {\frac{1} {N}}} \sum_{i=1}^{\left(M=129\right)} \sum_{j=1}^{\left(N= \mathrm{batch \, size}\right)}(O_{ij} - O^{emul}_{ij})^2
 \end{equation}
 \end{linenomath*}

The KL loss term can be interpreted as a regularizer of the resulting latent distributions \cite{Kingma2014}, which penalizes the complexity in the latent space based on the KL divergence.

\begin{linenomath*}
 \begin{equation}
 \mathrm{KL} \  \mathrm{loss} = {{ \frac{1} {2}} \times {\frac{1} {N}}} \sum_{j=1}^{\left(N=\mathrm{batch \,  size}\right)} \sum_{k=1}^{\left(K= \mathrm{latent \, space \, width}\right)}\left[-1 - \ln \sigma^2_{jk} + \mu^2_{jk} + \sigma^2_{jk}\right]
 \end{equation}
 \end{linenomath*}

 \begin{linenomath*}
 \begin{equation}
 \lambda \ \epsilon \ \mathbb{R_+} 
 \end{equation}
 \end{linenomath*}

We apply a KL annealing approach that multiplies the KL loss term by an annealing factor $\mathbf{\lambda}$ with initial value 0. The annealing factor then grows after a certain epoch during the training process \cite{Alemi2018}. This generally improves the reproduction capabilities of VAEs due to lowering the impact of the regularizing KL term \cite{Mooers2020}, avoiding a posterior collapse \cite{Alemi2018}, which negatively affects training. During a training step a 2D batch (dimensions $714 \times 64$) of 714 samples, the batch size, is fed into the VED to optimize the weights and biases. We use Adam as the VED’s optimizer\cite{Kingma2014Adam}. The purpose of an optimizer is to improve the networks performance (minimization of the networks loss function in our case) during the training process based on stochastic gradient descent. We choose this particular optimizer to follow the same strategy like in the preceding study of \citeA{Rasp2018}. The learning rate (the applied down-gradient step to optimize the loss) has an initial value of 0.00074594 based on a formal hyperparameter tuning and is divided by factor 5 after every 7$^{th}$ epoch over the course of the training. The batch size and the initial learning rate were chosen based on a formal hyperparameter search. Further optimized hyperparameters and a description of the hyperparameter search can be found in Table S1 and section S.1 of the supporting material. The chosen hyperparameters represent a suitable local minimum for the optimization of the VED architecture but should not be considered as the optimal hyperparameter setting.

\subsection{Benchmarking}

To benchmark the performance of our VED, we construct three reference networks with different architectures. The first reference network is an Encoder Decoder (ED). The ED closely mirrors the architecture of the VED except that there is no KL regularization, meaning that the calculation of  $\mathbf{\ln \sigma^2}$ and $\mathbf{\mu}$ is omitted. Furthermore, the ED's loss function only relies on the reconstruction loss. The second reference network, LR, is a further simplification of the ED, for which linear activations are used, which can be viewed as an equivalent to a principal component regression except that the latent space is not orthogonal. That is, the LR network can be interpreted as the combination of linear dimensionality reduction and regression modules. We use a reference deep artificial neural net (reference ANN) with its original output normalization based on \citeA{Rasp2018}, which was proven to be a skilful emulator of SPCAM. Note that to reproduce \citeA{Rasp2018}, meridional wind profiles were used as input fields to construct and train the reference ANN network. As an additional baseline model, we implement a linear version of our reference ANN. Similar to the reference ANN, this ``Reference Linear Model'' uses  256 nodes and 9 hidden layers but replaces all of the ANN's activation functions with the identity function (i.e. passing the values unchanged).
Finally, we constructed one further VED structure and a conditional VAE in the run-up of this study, which are presented in the supporting material (see section S.3) together with their strengths and limitations. Our goal is to strike a balance between the successful emulation of the target sub-grid-scale output data \textbf{Y} with compression, and the usefulness of scientific interpretation for convective processes and large-scale climate states. The VED we have chosen (see Figure \ref{fig:VAE_schematic}) is optimal on these fronts.

We split the SPCAM simulation into space-time shuffled training, unshuffled validation and unshuffled test data sets spanning 3 months ($\sim$ 4400 time steps) each. The input data \textbf{X} is normalised by subtracting the mean of each variable at each vertical level and dividing by the range between minimum and maximum of the resulting anomalies. Furthermore, we normalize the output of the VED, ED and LR as described in the supporting material (see section S.1). The output normalization, i.e., scaling to the same order of magnitude, allows us to achieve comparable reproduction skills across all fields. We show the impact of the existing differences of the VED output normalization and the reference ANN output normalization \cite{Rasp2018} on the evaluation of mean reproduction skills of the networks in section S.2 in the supporting material.  

In the next section we will evaluate the performance of the VED with respect to common reproduction metrics, and discuss the interpretability of the information encapsulated in the latent space.

\section{Evaluation of the VED}

In this section, we assess the predictive skill of the VED, and compare its mean regimes / statistics and tropical variability against reference networks. Furthermore, we evaluate the interpretability of the VED's latent space with respect to climate and convective variables. With this analysis, we are investigating the overall decoding (reproduction) and encoding (dimensionality reduction, interpretability) abilities of the VED to learn convective processes. 

\subsection{Mean Regimes and Statistics}

We start by evaluating the accuracy of the VED predictions to assess the impact of its dimensionality reduction on the overall performance. We use the mean squared error (MSE) to assess the performance of the VED predictions across sub-grid-scale fields \textbf{Y} for the training, validation, and test sets based on our VED output normalization. Overall, the VED shows good reproduction skills (see Table S4). The VED (test MSE = 0.165) clearly outperforms the linear model LR (test MSE = 0.243) in all data sets. The difference in predictive skills between VED and ED (test MSE = 0.165) is negligible. However, both networks express increased but comparable MSE with respect to reference ANN (test MSE= 0.135), in spite of the reference ANN having a substantially larger dimensionality (no latent manifold with a dramatic dimensionality reduction down to 5 nodes). These results are robust to the choice of output normalization \cite<VED's versus reference ANN's,>{Rasp2018}, as demonstrated in the supplemental material S.2.

\begin{figure}
    \centering
    \includegraphics[width=10.7cm]{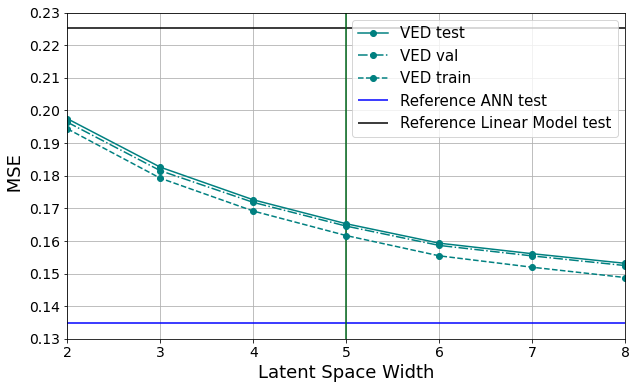}
    
    \caption{Mean squared error as a function of Latent Space Width of the VED for test (solid cyan), validation (dashed-dotted cyan) and training data set (dashed cyan curve) using our VED output normalization. The horizontal solid blue / black line represents the MSE scores of the reference ANN of \citeA{Rasp2018} / a linear version of this network (Reference Linear Model) on test data with fixed layer width of 256 nodes in the 9 hidden layers.}
    \label{fig:BN_dependence_combined}
    
\end{figure}

In the following, we explore whether a latent space of 5 nodes is a good compromise between accuracy to reproduce convective processes and physical interpretability in the latent space. Figure \ref{fig:BN_dependence_combined} shows the VED performance (MSE) on test, validation, and training data as a function of the latent space width. We find a substantial sensitivity of the VED’s performance to the latent space width - smaller width results in reduced accuracy associated with increased dimensionality reduction. Even for a latent space of two nodes, the VED has a higher predictive skill than the reference linear model, confirming the necessity of using nonlinear models to faithfully represent sub-grid-scale processes. Moreover, the VED's performance is converging towards the reference ANN for larger latent space widths (8 nodes). A latent space of 5 nodes results in a small reduction of predictive skills compared to the ‘wider’ latent space (Figure \ref{fig:BN_dependence_combined}), indicated by a MSE decrease of only $\approx$ 0.012 between a latent space of 5 nodes and 8 nodes. Additionally, we will show later (in Section 4) that such a latent space width enables the characterisation of realistic convective regimes and drivers of convective processes on specific nodes. This suggests that the overlap between different nodes is small. Despite this small overlap, we will show in section 4 that the resulting five latent nodes govern both SP convective processes and CAM climate states in most cases. For larger latent space widths of 6 nodes and more, the interpretability of resulting convective regimes gets more challenging due to the decaying impact of one latent node, or increasingly concurring influences between the nodes on SP convective processes or CAM climate variables. To summarize, regardless of how the output data are normalized (see Figure S1), the VED performs better than the reference linear model and approaches the performance of the fully-connected reference ANN as the latent space width increases.

As a complementary metric to evaluate the performance of the VED, we use the coefficient of determination \textbf{R$^2$} (Equation 5).

\begin{linenomath*}
\begin{equation}
    \mathbf{R^2}= 1 - \frac {\mathbf{MSE}}{\mathbf{Var}}
\end{equation}

\end{linenomath*}

\begin{linenomath*}
\begin{equation}
    \mathbf{MSE}= \frac{1}{\mathbf{P}} \sum_{t=1}^{ \mathbf{P}}(\mathbf{Y_{t}} - \mathbf{Y^{emul}_{t}})^2
\end{equation}

\end{linenomath*}

\begin{linenomath*}
\begin{equation}
    \mathbf{Var}= \frac{1}{\mathbf{P}} \sum_{t=1}^{ \mathbf{P}}(\mathbf{Y_{t}} - \frac{1}{\mathbf{P}} \sum_{t=1}^{ \mathbf{P}}\mathbf{Y_{t}})^2
\end{equation}

\end{linenomath*}

\begin{figure}[ht]
    \centering 
    \includegraphics[width=13.8cm]{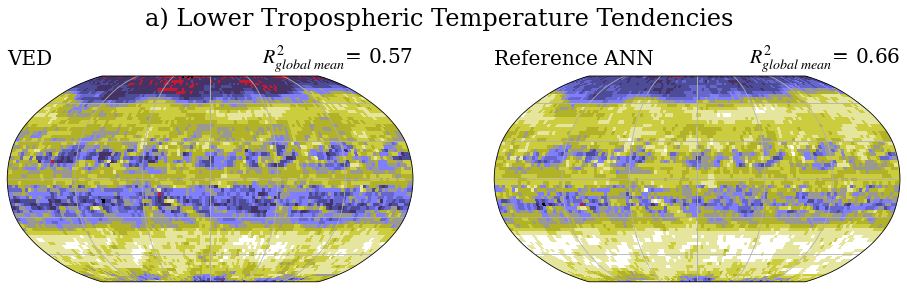}

    \caption{Coefficient of determination \textbf{R$^2$} of lower tropospheric temperature tendencies (a) and lower tropospheric specific humidity tendencies (b) at 700 hPa for the VED (left) and reference ANN (right column).  The global mean \textbf{R$^2$} of each field is indicated in the upper right above every subplot.}
    \label{fig:r_2_scores}
\end{figure}

It is defined as the difference of 1 and the ratio between the mean squared error (\textbf{MSE}) and the true variance (\textbf{Var}) of the data, where \textbf{P} is the length of the time series, \textbf{t} is the respective time step and \textbf{Y} / \textbf{Y$^{emul}$} are the true value of the test data / VED prediction.
We constructed at first the time series of all output variables \textbf{O} from the test data set or predictions and computed the respective coefficients of determination in each grid cell ( 64 points in latitude $\times$ 128 points in longitude = 8192) of all layers. We selected the global sub-grid heating and moistening fields at 700 hPa for the evaluation of the VED's \textbf{R$^2$} (Figure \ref{fig:r_2_scores}).

We choose \textbf{dq/dt} and \textbf{dT/dt} fields at this pressure level because of the limited skill in fitting lower tropospheric convective processes with neural nets that has been reported across multiple investigations, and which has been speculated to be associated with an underrepresentation of stochastic variability linked to shallow and deep convection \cite{Gentine2018,Rasp2018,Mooers2021,Wang2021}. Both networks, VED and reference ANN, exhibit similar emulation skill patterns for heating and moistening tendencies, including the skill deficits for low-level moistening tendencies in the tropics, as seen in previous studies. Overall, we see a decreased reproduced variability with the VED (\textbf{R}$^2_{\mathrm{global \, mean}}$ = 0.57 / 0.42 for \textbf{dT/dt} / \textbf{dq/dt}; 35$\%$ and 22$\%$ of horizontal grid cells for temperature and specific humidity tendencies with \textbf{R}$^2$ $>$ 0.7, respectively) compared to the reference ANN (\textbf{R}$^2_{\mathrm{global \, mean}}$ = 0.66, 0.53 for \textbf{dT/dt}, \textbf{dq/dt}; 51$\%$ and 35$\%$ of horizontal grid cells with \textbf{R$^2$} $>$ 0.7 for temperature and specific humidity tendencies, respectively). The VED shows regions of high reproduction skill for both, temperature and specific humidity tendencies along the mid-latitude storm tracks ($\sim$ 45$^{\circ}$ N / S, \textbf{R$^2$} $\sim$ 0.7) and in the Inter-Tropical Convergence Zone (ITCZ) region near the equator (ascending branch of Hadley Cell associated with deep convection, \textbf{R$^2$} $\sim$ 0.6). Both networks exhibit weaker prediction skill of specific humidity and temperature tendencies near the descending branches of the Hadley Cell (subtropical highs $\sim$ 20$^{\circ}$ N / S) associated with an underestimation of (shallow) convective variability. \citeA{Mooers2021} also found comparably weaker reproduction skill of their neural net in this region. Recently \citeA{Wang2022} showed that the reproduction of moistening tendencies in the subtropics can be improved by using non-local features from adjacent grid cells as additional inputs of the neural net.
Nevertheless, the VED shows good reproduction skill associated with convective processes in the lower troposphere compared to the reference ANN, despite its strongly reduced dimensionality in the latent space. This suggests that the information from large-scale climate variables \textbf{X} that is relevant for the prediction of sub-grid-scale convective processes \textbf{Y} is closer to 5 (our latent space's dimensionality) than 64 (the input vector length). In other words, this means that the number of large-scale variables needed to skillfully emulate sub-grid-scale processes is far smaller than the number of original input variables of the superparameterization. This is consistent with assumptions made by reduced-complexity models, such as the lower-dimensional multi-cloud model \cite{Frenkel2012} or the quasi-equilibrium tropical circulation model \cite{Neelin2000}.  

\subsection{Tropical Variability}

Current ESMs exhibit large biases in tropical precipitation and associated patterns \cite{Bock2020}. These regional uncertainties can be attributed to the fact that many ESMs struggle to reproduce tropical intra-seasonal variability like the Madden Julian Oscillation \cite<MJO, an eastward propagating pattern of clustered deep convection in the Indo-Pacific Region;>{Zhang2005}. SPCAM yields a more realistic reproduction of the MJO compared to the traditional convective parametrization of CAM \cite{Khairoutdinov2005}. Furthermore, the governing tropical variability is largely reproducible with deep learning approaches \cite{Rasp2018}. Here, we investigate the ability of the VED to not distort the high-frequency tropical variability (15$^\circ$ N to 15$^\circ$ S) as simulated by SPCAM compared to the reference ANN. For this analysis, we use the entire second year of the SPCAM simulation to identify driving tropical variability with frequency lower than $\frac{1}{30} \ \mathrm {days^{-1}}$. This second SP year includes the 3-month sequence of the validation data set but has no overlap with the training data set.  

\begin{figure}
    \centering
    \includegraphics[width=13.5cm]{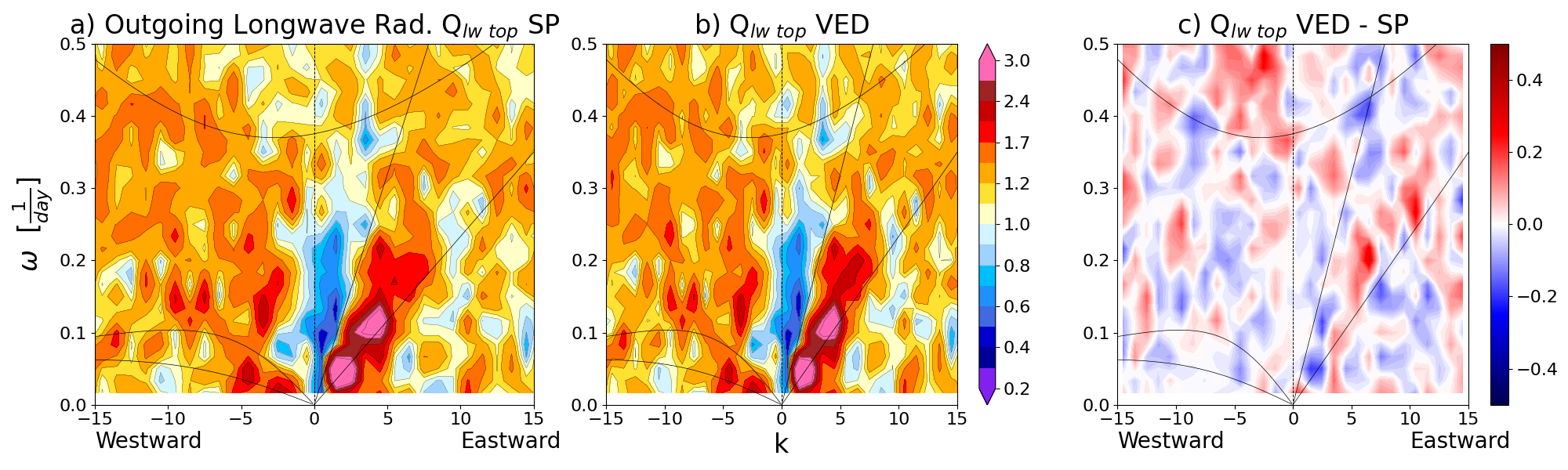}
    \caption{Wheeler Kiladis diagram based on tropical outgoing longwave radiation [15°N-15°S] of SP (a), of VED predictions (b) and the absolute difference of spatio-temporal wave spectra VED-SP (c) for 1 year of SP simulations}
    \label{fig:WK_VAE_clim_clim_conv}
\end{figure}

Figure \ref{fig:WK_VAE_clim_clim_conv} shows the Wheeler-Kiladis diagrams, diagnosing the equatorial symmetric component (zonal wave numbers \textbf{k}) of outgoing longwave radiation (\textbf{Q$_{lw \ top}$}) with respect to its frequency $\mathbf{\omega}$ for both SPCAM (Figure \ref{fig:WK_VAE_clim_clim_conv}a) and VED (Figure \ref{fig:WK_VAE_clim_clim_conv}b). Eastward propagating, non-dispersive Kelvin waves ($\mathbf{\omega^{-1}}$ $\sim$ 8 - 20 days, \textbf{k} $\sim$ 2-5) and the MJO ($\mathbf{\omega^{-1}}$ $\sim$ 30 days, \textbf{k}=1) are not distorted by the VED. The resulting differences in the reproduced spatio-temporal variability with respect to SPCAM are generally confined within -0.2 to 0.2 (unit-less values) (Figure \ref{fig:WK_VAE_clim_clim_conv}c), which amounts to a relative error of roughly 20$\%$, and are not associated with a damping or absence of general features in $\mathbf{\omega}$-\textbf{k} space. 

Although the reference ANN shows slightly better reproduction skill (see Figure S3 in the supporting material), the VED and also ED (see Figure S2) can realistically reproduce not only mean regimes and characteristics of convective processes but also the associated variability even with its strongly reduced dimensionality on only 5 latent nodes. 

Next, we evaluate our main interest -- the physical interpretability of the VED with respect to convective processes -- by exploring the information encapsulated in its latent space. We will show in the following sections that the representation of general convective processes is actually much lower dimensional than potentially envisioned.

\subsection{Interpretability via Latent Space Exploration} 

In this section, we investigate convective processes and large-scale climate states captured in the latent space of the VED. This will give us a first impression of general drivers of convective predictability encapsulated in the latent manifold and will show the potential to study convective processes with only five latent nodes. Latent spaces of VAEs behave to some extent as a non-linear equivalent of a principal component analysis, PCA \cite<e.g.,>{Rolinek2019}, due to a skilful lower-dimensional encoding of information fed into the network. Therefore, we test whether the latent space of the VED retains a meaningful lower dimensional representation of convective processes like we would expect from a traditional PCA.

\begin{figure}[ht]
    \centering
    \includegraphics[width=13.9cm]{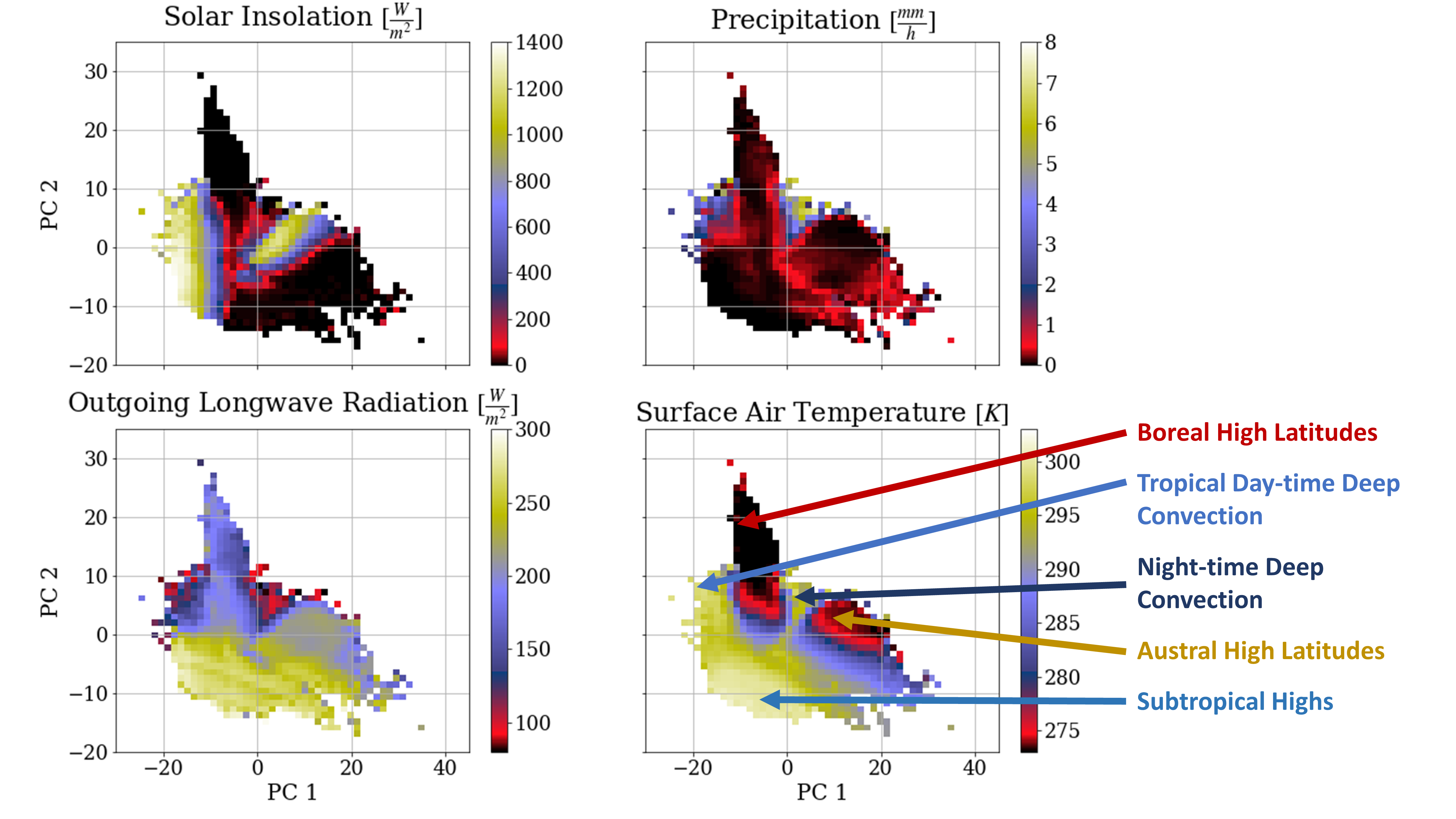}
   
    \caption{The 2D PCA-compressed latent space of the VED and associated conditional averages of solar insolation (upper left), precipitation (upper right), outgoing longwave radiation (lower left) and surface air temperature (lower right panel) of projected SP test data (see color scheme in each subplot). The x-axis / y-axis in all subplots indicates the 1$^{st}$ / 2$^{nd}$ leading PC of the 5D latent space, which have a combined ``explained variance'' of around 0.82. The arrows in the lower right subplot indicate the position of characteristic samples from different geographic regions inside the 2D PCA-compressed latent space of the VED mentioned in the text.}
    \label{fig:VAE_clim_clim_conv_glob_multi}
\end{figure}

Human visualization of the full five latent dimensions (5 nodes, 5D) in a 2D schematic requires some additional dimensionality reduction. For visualization purposes, we therefore use a PCA to first compress the 5D manifold into a 2D lower-dimensional embedded space, which allows a visual inspection of the encapsulated information. The resulting 2D PCA representation contains 82$\%$ of the total variance of the VED’s latent space. Figure \ref{fig:VAE_clim_clim_conv_glob_multi} shows the first (x-axis) and second leading PC (y-axis) of the compressed latent space for 1 million randomly sampled points. The manifold, which is spanned by the two leading PCs, is then divided into a regular grid of size 50 (PC 1) $\times$ 50 (PC 2) cells. Tracking each selected sample allows us to characterize the embedded information for both convection and large-scale climate states. This permits us to compute conditional averages of these convection related variables in each grid cell of the 2D PCA compressed manifold.  

Figure \ref{fig:VAE_clim_clim_conv_glob_multi} depicts the conditional averages of solar insolation (\textbf{Q$_{sol}$}), precipitation (\textbf{precip}), outgoing longwave radiation (\textbf{Q$_{lw \ top}$}), and surface air temperature (\textbf{T$_{surf}$}) in the 2D PCA compressed latent space of the VED. Together the results show that distinct convective regimes are clearly separated in the latent space. More information on how the complex global superposition of distinct geographic convective regimes and large-scale processes in the latent space is contributed by separate latitudinal bands of the aquaplanet (tropics, boreal and austral mid latitudes) is provided in the supporting information. Therein Figure S4 shows the fixed SSTs of the simulation and S6 the regional decomposition of patterns in the VED's latent space. These two figures can aid as a reference guide for the following latent space exploration. We start the analysis by investigating the impact of the insolation \textbf{Q$_{sol}$} on the latent space position, including whether the expected diurnal cycle of convective processes in SPCAM \cite{Khairoutdinov2005,Pritchard2009} is manifested in the latent space of the VED. Indeed, solar insolation \textbf{Q$_{sol}$} reveals 2 distinct maxima with day-time conditions and 2 minima with night-time conditions, which are separated by diurnal transition zones, as expected from diurnally varying input and output data of SP. Cross-evaluating the conditional averages of solar insolation with \textbf{T$_{surf}$}, one can diagnose that the 2D PCA compressed latent space of VED stores information that can be used to infer the geographic location of a sample. As an example, we can focus on the ‘fin-shaped’ region (PC1 $\sim$ -8, PC2 $\sim$ 15) protruding from the top of the 2D PCA compressed latent space. Here the samples are characterized by anomalously cold (273 K $<$ \textbf{T$_{surf}$} $<$ 278 K) climate conditions without solar insolation. Based on the fixed SST forcing (see Figure S4), the low surface air temperatures and the constant perpetual Austral Summer solar forcing, we can conclude that these samples originate from polar and subpolar latitudes in the Northern Hemisphere. Furthermore, we find a zone with day-time solar insolation (\textbf{Q$_{sol}$} $>$ 700 $\mathrm{\frac{W}{m^2}}$) and cold surface air temperature (273 K $<$ \textbf{T$_{surf}$} $<$ 280 K) in the upper-right part of the latent space (PC1 $\sim$ 10, PC2 $\sim$ 5), which represents large-scale climate conditions that can be only found in the austral polar and subpolar latitudes in SPCAM test data. 

We also explore the fingerprinting of precipitation on the latent space as a proxy for the strength of convective processes, since it is closely connected to convective moistening and convective heating \cite{Emanuel1994,Lohmann2016}. The 2D PCA compressed latent space of the VED reveals a good separation of samples with no or negligible precipitation, shallow convection with the formation of weak precipitation, and deep convective samples with intense precipitation (\textbf{precip} $>$ 10 $\frac {mm}{h}$ in the tropics, see Figure S6). We expect to see a clear separation between tropical deep convective samples and samples with no or negligible precipitation from the colder higher latitudes or the region of the subtropical highs in the 2D PCA compressed latent space due to the strong variation in the magnitude of convective processes with latitude as it is visible in Figure \ref{fig:VAE_clim_clim_conv_glob_multi}. If we now focus on the conditionally-averaged plot of precipitation, two maxima are evident. The first precipitation maximum (PC1 $\sim$ -15, PC2 $\sim$ 5) is associated with day-time solar forcing, a minimum of outgoing long-wave radiation (\textbf{Q$_{lw \ top}$} $<$ 150 $\frac{W}{m^2}$, which suggests high cloud tops in the upper half of the troposphere) and tropical surface air temperatures (\textbf{T$_{surf}$} $>$ 295 K). Therefore, this maximum originates from tropical day-time deep-convective samples in SPCAM. The second maximum (PC1 $\sim$ 5, PC2 $\sim$ 5) exhibits slightly colder surface air temperatures, night-time conditions, decreased outgoing longwave radiation (\textbf{Q$_{lw \ top}$} $\sim$ 100 $\frac{W}{m^2}$) and precipitation formation of more than 3 $\frac{mm}{h}$. It can be shown that this maximum originates from night-time deep convection from the tropics in its center and predominantly strong precipitating samples from the Northern and Southern extratropics along the left and right boundary, respectively. 

Outgoing longwave radiation (\textbf{Q$_{lw \ top}$}) is a good estimator for both the height of cloud tops based on the inferred brightness temperatures for convective samples or surface temperatures for non - or negligibly - convective samples. Based on the combination of high \textbf{Q$_{lw \ top}$} (no or negligible convection), no precipitation formation and anomalous warm surface temperatures (\textbf{T$_{surf}$} $\sim$ 300 K), one can conclude that samples from subtropical highs (the descending branch of the Hadley cell, with limited deep-convective processes with large vertical extent in the free troposphere) are concentrated in the lower left part of the PCA compressed latent space (PC1 $\sim$  -10, PC2 $\sim$ -10) of the VED. 

These results demonstrate how large-scale climate conditions and convective processes are connected and physically interpretable in the latent space (e.g., equivalence of \textbf{precip} maxima and \textbf{Q$_{lw \ top}$} minima), which illustrates the encoding power of the VED. Furthermore, the evaluated mean statistics support that the VED realistically reproduces convective processes and the associated variability despite a strong dimensionality reduction down to only five nodes in the latent space, which shows the decoding power of the network. 

Similar reproduction abilities can be investigated for ED, but the physical interpretability of the resulting latent space is reduced compared to VED. The KL divergence used for the VED ensures an improved separation of latent modes. The effect can be seen in a larger number of centers of action in the ED’s latent space and weaker gradients in the conditional average plots with respect to sub-grid-scale and climate variables, as can be seen in the supporting material Figure S5 (AE vs VED latent spaces) and S6 for the VED conditional average plot or S7 for ED conditional average plot. Additionally, we tested the interpretability of the 2D PCA compressed latent space of a VED trained on \textbf{X} to emulate \textbf{Y}, in other words mirroring the input data and output data of SP (see S.3A in the supporting material). In this case the latent space strongly focuses on the magnitude of heating or moistening tendencies, resembling a weak gradient from negligible convective processes towards strongly precipitating deep convective samples (see Figure S16). For large-scale climate variables like surface air temperature, the 2D PCA compressed latent space of a VED$_{X \rightarrow Y}$ mostly distinguishes between warm conditions and cold conditions sorting samples from both poles close together in one minimum (see Figure S17), which makes the visual separation of austral and boreal polar latitudes nearly impossible. In contrast, VED shows a pronounced separation of austral and boreal polar samples and reveals distinct regimes of convective processes in its 2D PCA compressed latent space as seen in Figure \ref{fig:VAE_clim_clim_conv_glob_multi}, which is a clear advantage in interpretability of this network compared to VED$_{X \rightarrow Y}$. 

We further compared the interpretability of the 2D PCA compressed latent space of the VED against a traditional PCA on the large-scale input features \textbf{X}, as an unsupervised linear reference method. The first two leading PC's with respect to \textbf{X} show overall weak gradients in its lower-dimensional space for the conditional averages of solar insolation, outgoing longwave radiation and surface air temperature (Figure S8). The `centers of action' are less pronounced for the PCA on \textbf{X} compared to its equivalent on the latent space of VED seen in Figure \ref{fig:VAE_clim_clim_conv_glob_multi}. Especially the identification of deep convective samples is hardly possible inside the submanifold spanned by the two leading PC's of the large-scale variables as can be seen in Figure S8. In latitude - longitude plots (Figure S9) these two leading PC's resemble large-scale patterns with meridional gradients that show similarities with temperature or radiation fields but barely with sub-grid-scale variables. In contrast, the latent space of VED focuses on both large-scale and sub-grid-scale patterns. The first two latent variables are characterised by large-scale patterns connected to geographic variability and solar insolation (see Figure S10). The remaining three latent variables describe mostly sub-grid-scale convective processes, as can be seen in Figure S10.\\
The concept of the computing conditional averages can be repeated also on 2D projections spanned by the original latent variables of the VED without a PCA as postprocessing step. An example of this more detailed latent space inspection can be found in the supplementary material in section S.2 (Figure S12 for precipitation, S13 for solar insolation and S14 for surface air temperature).

As a next step, we will combine the reproduction skill and the encapsulated information in the latent space of the VED to investigate convective processes by identifying distinct large-scale drivers, associated convective regimes and geographic variability in detail. 

\section{Unveiling Drivers of Convective Processes in SPCAM Using Generative Modeling}

In this section, we discuss the dominant drivers of convective processes encapsulated in the latent space of the VED using a generative modeling approach. We compute the marginal distributions of all 5 latent variables \textbf{z}. We focus on the 10$^{th}$, 25$^{th}$, 50$^{th}$, 75$^{th}$ and 90$^{th}$ percentiles of the marginal distributions of the latent variables. Since most of these distributions are bi-modal (see Figures \ref{fig:gen_Node_1} – \ref{fig:gen_Node_5}a), we select their median values as estimators for the intersect (origin) of the 5-dimensional \textbf{z}, instead of the mean. For all latent variables, the median is close to the mode value (peak value) of the marginal distributions. To generate the `median' climate conditions and associated convective processes from the `median' values of the latent variables, we construct a reference state \textbf{z$_{median}$} (Equation 8). \textbf{z$_{median}$} contains the median values for all five latent variables. This reference state is fed into the decoder of the VED to generate vertical heating, moistening, specific humidity, and temperature profiles. These vertical profiles represent the `median' state of convective processes and associated climate conditions.

\begin{linenomath*}
\begin{equation}
    \mathbf{z_{median}}=\left[\mathrm{median\left(z_1\right), median\left(z_2\right), median\left(z_3\right) ,median\left(z_4\right) ,median\left(z_5\right)}\right] 
\end{equation}

\end{linenomath*}

To investigate encapsulated convective regimes and large-scale climate states in the latent space of VED via generative modeling, we replace the median value with the different percentiles (perc$\mathrm{\,(z_1)}$ in Equation 9) along one specific marginal distribution. This analysis identifies how each latent node drives a variation of convective processes and large-scale climate states generated by the decoder and manifests in well-known convective regimes. The modified \textbf{z$_{translation}$} (Equation 9) can be seen as a latent forcing on the decoder, acting as a knob which amplifies or damps the associated convective features. Furthermore, \textbf{z$_{translation}$} influences the geographic variability of generated samples, allowing an interpolation from a tropical to a polar `background' climate state like a knob for the general volume of generated large-scale profiles. A clear separation between geographic versus convective modulation with a distinct \textbf{z$_{translation}$} is challenging and not the primary goal of our VED's decoder setup. The evaluation whether a distinct latent node drives more geographic than convective modulation necessarily involves an analysis of all generated variables - an interesting analysis trade-off revealed by this latent space exploration.  \textbf{z$_{translation}$} can be geometrically interpreted as a translation along one distinct latent dimension in the 5-dimensional latent space. For instance, \textbf{z$_{translation}$} is applied as an example to latent node 1 perturbing the ‘median’ conditions along this latent dimension, while keeping the median values for the 4 other dimensions: 

\begin{linenomath*}
\begin{equation}
    \mathbf{z_{translation \, node \, 1}} = \left[\mathrm{perc\left(z_1\right), median\left(z_2\right), median\left(z_3\right) ,median\left(z_4\right) ,median\left(z_5\right)}\right]
\end{equation}

\end{linenomath*}

Applying a translation along one latent dimension while keeping the other latent variables fixed to their median values implicitly assumes that latent variables do not overly depend on each other. To test this independence, we calculate the Pearson correlation between all five latent variables using the entire test data set. The mean correlation coefficients between the latent dimensions are confined within $\pm 0.35$, except for a mean correlation of -0.74 between latent variables 2 and 5. The relatively large linear connection between latent variables 2 and 5 can be further explored by density plots using the 2D projection spanned by these latent variables, see Figure S11. While Latent Node 2 separates moist and warm from cold and dry tropospheric conditions, Latent Node 5 represents deep convection samples, which rely on anomalous wet and warm conditions in the troposphere. Therefore it is not surprising to see a pronounced anti-correlation between these nodes. This is a further evidence of the interpretability and meaningfulness of the VED's latent space, i.e., learning physical processes in the lower-order manifold.

In the following we will use \textbf{z$_{translation}$} along all five latent dimensions to identify large-scale drivers of convective processes and different convective regimes in SPCAM. We use the notation `high \textbf{z$_{translation}$}' to describe the cases when \textbf{z$_{translation}$} $\color{blue}{>}$ \textbf{z$_{median}$} and `low \textbf{z$_{translation}$}' if \textbf{z$_{translation}$} $\color{blue}{<}$ \textbf{z$_{median}$}. Figures \ref{fig:gen_Node_1} – \ref{fig:gen_Node_5} illustrate the marginal distribution along the respective latent dimensions (Panels a, where the dashed black line indicates the median value of each dimension, Equation 8). The other subplots of these figures show the generated vertical moistening, heating, specific humidity and temperature profiles (Panels b-e) of the decoder with respect to \textbf{z$_{median}$} (Equation 8) or \textbf{z$_{translation}$} (Equation 9 along a distinct latent dimension). Additionally, two sub-grid-scale and climate variables (Panels f), which are strongly affected by the applied latent forcing, are displayed as a function of \textbf{z$_{translation}$} for illustrative purposes. The marker-edge-color in the respective Panels f reveal the chosen percentiles. All other generated sub-grid-scale and large-scale climate variables are shown in Tables S7-11 in the supporting material. We investigate in the following that latent node 1 and latent node 2 focus on the large-scale climate (geographic) variability in \textbf{X} rather than on sub-grid-scale convective processes in \textbf{Y}. In contrast, latent nodes 3, 4 and 5 exhibit main characteristics of dominant convective regimes captured in \textbf{Y}.

\subsection{Large-Scale Climate Variability Nodes}

In this first part we demonstrate that latent nodes 1 and 2 capture mostly large-scale climate variability in \textbf{X}.

\subsubsection{Latent Node 1: Global Temperature Variations}

Global temperatures in the troposphere are dominated by the large meridional gradients from equatorial to polar latitudes mainly related to solar insolation differences between the tropics and extratropics. 

\begin{figure}[ht]
     \centering
     \includegraphics[width=13.5cm]{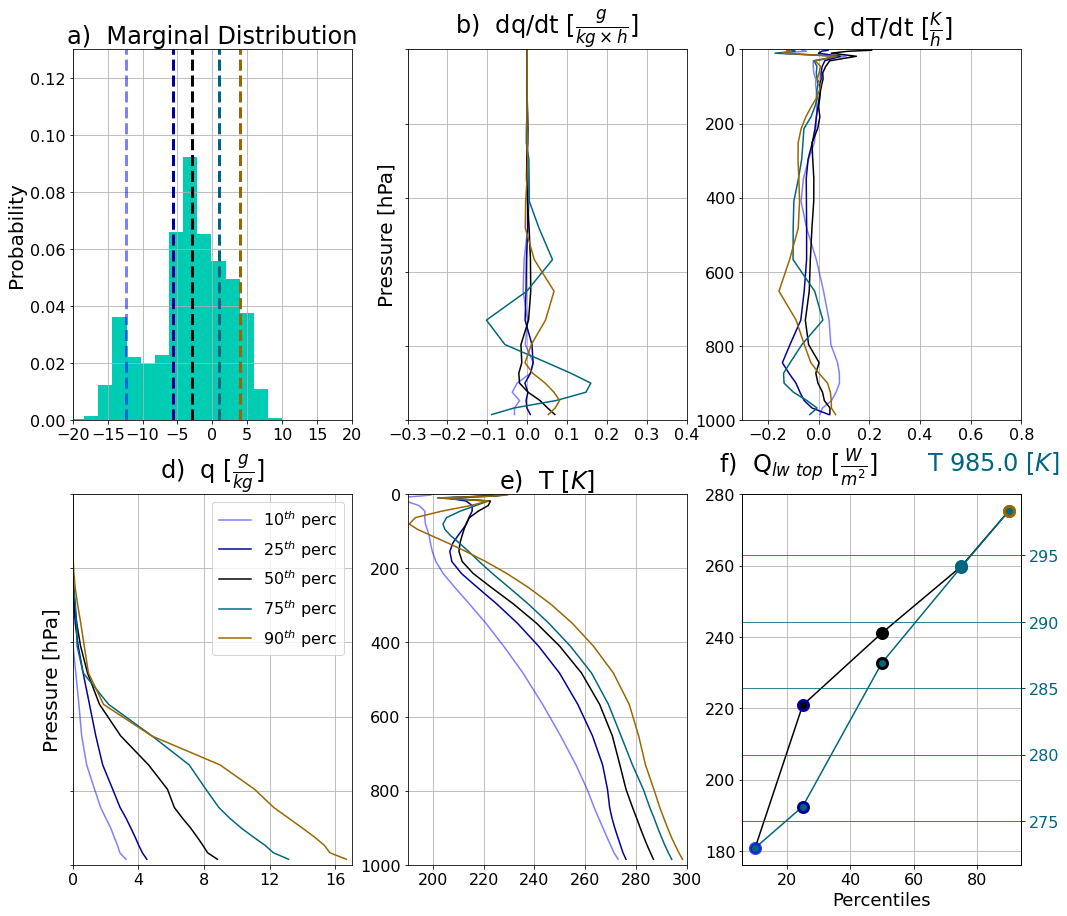}
     \caption{Marginal distribution of latent node 1 (a) and the resulting generated vertical profiles of specific humidity tendencies \textbf{dq/dt} (b), temperature tendencies \textbf{dT/dt} (c), specific humidity \textbf{q} (d) and temperatures \textbf{T} (e). The dashed lines in the marginal distribution plot represent the chosen percentiles (see legend in subplot d) and the resulting effect of the respective translation \textbf{z$_{translation}$} on the profiles is shown in the subplots. Furthermore, the longwave heat flux at the model top (\textbf{Q$_{lw \ top}$}) and the surface air temperature (\textbf{T$_{surf}$}  / \textbf{T 985.0}) (f) are illustrated as function of the translation \textbf{z$_{translation}$} along the latent dimension 1. The marker-edge-color in panel f symbolise the respective percentiles of \textbf{z$_{translation}$}}. The black lines in subplots b-e indicate the generated reference state with \textbf{z$_{median}$}.
     \label{fig:gen_Node_1}
 \end{figure}

The first latent node (Node 1) captures these global meridional temperature variations (Figure \ref{fig:gen_Node_1}e), as suggested by the large spread of the surface temperature response to \textbf{z$_{translation}$}, encompassing the tropics (\textbf{T$_{surf}$} $\sim$ 298 K, high \textbf{z$_{translation}$}) and polar regions (\textbf{T$_{surf}$} $\sim$ 273 K,  low \textbf{z$_{translation}$}). Tropical regions are characterized by very moist conditions in the boundary layer (\textbf{q} $>$ 10 $\frac{g}{kg}$), while being extremely dry at the poles (\textbf{q} $\sim$ 1.5 – 3.5 $\frac{g}{kg}$), see Figure \ref{fig:gen_Node_1}d. The strong connection between tropospheric temperatures or specific humidity and Node 1 can be shown with a linear correlation of globally concatenated temperature space-time series (of horizontal grid cells and time, featuring the large meridional gradients) and respective node space-time series. The resulting ``linear explained variance'' of temperature space-time series on Node 1 exceeds 0.5 (Figure S18), while the ``linear explained variance'' vanishes if the analysis is repeated on the time series for each horizontal grid cell (Figure S19, without the large meridional gradients). A detailed description how these two correlations metrics were computed can be found in the supporting material in section S.4.

A physical interpretation of this response on the \textbf{z$_{translation}$} can be given based on the Clausius-Clapeyron relationship. A warmer atmosphere results in a near-exponentially higher saturation water vapor pressure, which in turn allows higher specific humidity content. Therefore, we see strongly coupled variations of temperature and specific humidity between the equator and the poles. In short, the first latent node represents these overarching large-scale meridional variations in tropospheric temperatures, influencing specific humidity, but is not necessarily linked to convective processes \textbf{Y}, but rather to large-scale conditions \textbf{X}, which are also part of the VED reconstruction.

\subsubsection{Latent Node 2: Large-Scale Variability along the Mid-Latitude Storm Tracks}

\begin{figure}[ht]
     \centering
     \includegraphics[width=13.5cm]{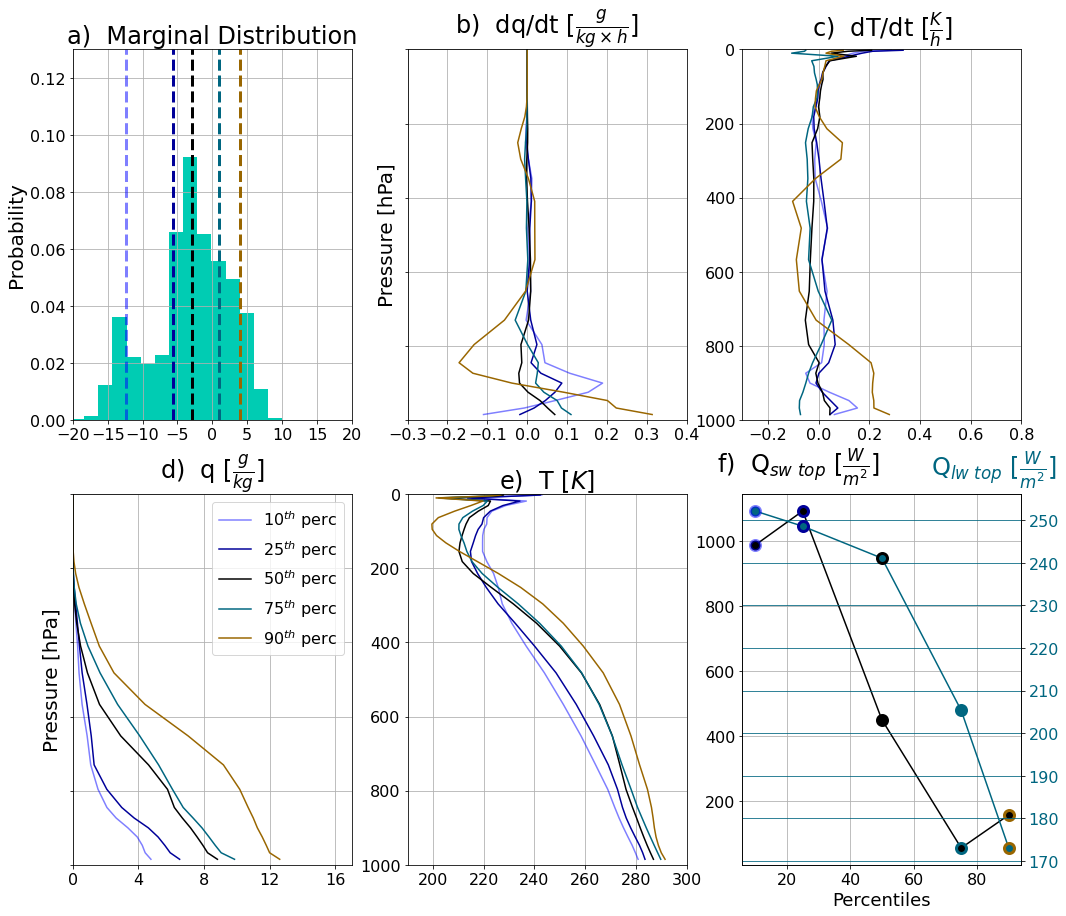}
     \caption{Marginal distribution of latent node 2 (a) and the resulting generated vertical profiles of specific humidity tendencies \textbf{dq/dt} (b), temperature tendencies \textbf{dT/dt} (c), specific humidity \textbf{q} (d) and temperatures \textbf{T} (e). The dashed lines in the marginal distribution plot represent the chosen percentiles (see legend in subplot d) and the resulting effect of the respective translation \textbf{z$_{translation}$} on the profiles is shown in the subplots. Furthermore, the shortwave heat flux at the model top (\textbf{Q$_{sw \ top}$}) and the outgoing longwave heat flux (\textbf{Q$_{lw \ top}$}) (f) are illustrated as function of the translation \textbf{z$_{translation}$} along the latent dimension 2. The marker-edge-color in panel f symbolise the respective percentiles of \textbf{z$_{translation}$}.} The black lines in subplots b-e indicate the generated reference state with \textbf{z$_{median}$}.
     \label{fig:gen_Node_2}
 \end{figure}

Latent node 2 characterizes more the large-scale climate (and thus geographic) variability in \textbf{X} than focuses on a distinct convective regime. Latent dimension 2 (Node 2, Figure \ref{fig:gen_Node_2}) clearly captures temperature and specific humidity variations in the troposphere, as can be seen in Figure \ref{fig:gen_Node_2}d and \ref{fig:gen_Node_2}e. Warmer and moister tropospheric conditions are associated with high \textbf{z$_{translation}$}. 
 
 Low \textbf{z$_{translation}$} characterizes cold and stable conditions during day-time (\textbf{Q$_{sw \ top}$} $\sim$ 1000 $\frac{W}{m^2}$). These anomalous cold and dry conditions in the upper troposphere are associated with negligible convective processes, as diagnosed with a large outgoing longwave heat flux at the model top (\textbf{Q$_{lw \ top}$} $\sim$ 240 $\frac{W}{m^2}$) and the formation of no precipitation (Table S8). Due to the large shortwave heat flux at the model top, the perpetual austral summer solar forcing and the low surface air temperatures (\textbf{T$_{surf}$} $\sim$ 281 K), low \textbf{z$_{translation}$} can be traced back to the austral mid-latitudes. Whereas high \textbf{z$_{translation}$} is linked to night-time conditions (\textbf{Q$_{sw \ top}$} $<$ 200 $\frac{W}{m^2}$) with a warm, moist troposphere (\textbf{T$_{surf}$} $\sim$ 291 K). High \textbf{z$_{translation}$} is further characterized by mid-level convection (\textbf{Q$_{lw \ top}$} $\sim$ 180 - 200 $\frac{W}{m^2}$) with intermediate precipitation formation (\textbf{precip} $\sim$ 0.12 to 0.15 $\frac{mm}{h}$, Table S8) associated with a warmer and moister upper troposphere and can be found in the subtropics on both hemispheres.
 
Our approach allows us to identify the main patterns of the large-scale climate state in \textbf{X}, which are main drivers of the general circulation and convection, besides convective processes in \textbf{Y} in the latent space. These convective processes are heavily modulated by \textbf{X}. Node 2 captures characterising features of the large-scale meridional variability of specific humidity and temperatures between the mid latitudes and the subtropics \cite<i.e., an essential driver of mid-latitude storm track dynamics;>{Bony2015}. Latent dimension 2 is further influenced by the solar forcing. The clear separation between austral mid latitude temperature profiles on one side and samples from subtropical regions on the other side of the \textbf{z$_{translation}$} are further evidence that latent node 2 encapsulates a part of the geographic variability inside the latent space seen in Figure \ref{fig:VAE_clim_clim_conv_glob_multi}.

 \subsection{Convective Regime Nodes}
 
Next, we will show that latent node 3, 4, 5 usefully characterize mostly distinct convective regimes in the sub-grid-scale process rate variables \textbf{Y}.

\subsubsection{Latent Node 3: Shallow Convection}

Shallow convective processes are one of the dominant cloud regimes investigated in observational studies \cite<e.g.,>{Huaman2017}. Latent node 3 characterizes some of the main characteristics of shallow convective processes as revealed by its vertical profiles of specific humidity and temperature tendencies influenced by large-scale specific humidity and surface diabatic fluxes.

\begin{figure}[ht]
     \centering
     \includegraphics[width=13.5cm]{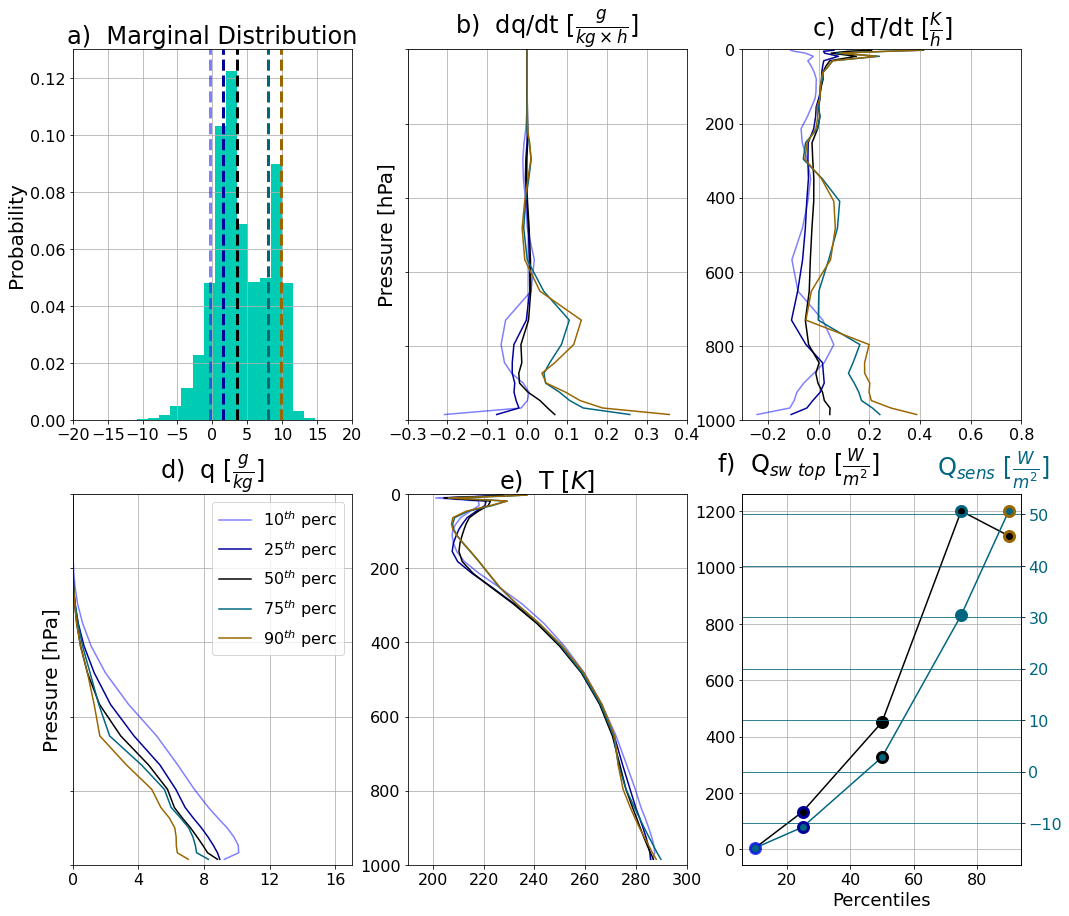}
     \caption{Marginal distribution of latent node 3 (a) and the resulting generated vertical profiles of specific humidity tendencies \textbf{dq/dt} (b), temperature tendencies \textbf{dT/dt} (c), specific humidity \textbf{q} (d) and temperatures \textbf{T} (e). The dashed lines in the marginal distribution plot represent the chosen percentiles (see legend in subplot d) and the resulting effect of the respective translation \textbf{z$_{translation}$} on the profiles is shown in the subplots. Furthermore, the shortwave heat flux at the model top (\textbf{Q$_{sw \ top}$}) and the surface sensible heat flux (\textbf{Q$_{sens}$}) (f) are illustrated as function of the translation \textbf{z$_{translation}$} along the latent dimension 3.  The marker-edge-color in panel f symbolise the respective percentiles of \textbf{z$_{translation}$}. The black lines in subplots b-e indicate the generated reference state with
     \textbf{z$_{median}$}.}
     \label{fig:gen_Node_3}
 \end{figure}

Figure \ref{fig:gen_Node_3} shows the marginal distribution of latent node 3 (Node 3), the generated vertical specific humidity and temperature tendencies, and the large-scale specific humidity and temperature profiles of the Decoder for \textbf{z$_{median}$}, as well as the applied \textbf{z$_{translation}$}. Furthermore, the generated shortwave heat flux (\textbf{Q$_{sw \ top}$}) and surface sensible heat flux (\textbf{Q$_{sens}$}) are displayed as a function of \textbf{z$_{translation}$}. Along latent dimension 3, the specific humidity (\textbf{q}) decreases throughout the entire troposphere for increasing \textbf{z$_{translation}$}, while surface diabatic fluxes (sensible heat flux \textbf{Q$_{sens}$} and latent heat flux \textbf{Q$_{lat}$}, Table S9) increase. Likewise, the outgoing longwave radiation \textbf{Q$_{lw \ top}$} increases with increasing \textbf{z$_{translation}$} suggesting higher cloud tops and stronger convective processes for low \textbf{z$_{translation}$} (Table S9). In contrast, the intensity of shallow convection and outgoing longwave radiation decreases when \textbf{z$_{translation}$} increases (high \textbf{z$_{translation}$}). Specific humidity tendencies (\textbf{dq/dt}) in the lower troposphere (p $>$ 600 hPa) react to \textbf{z$_{translation}$} in a bimodal way. They moisten, in combination with a strong positive surface diabatic forcing, the relatively dry ambient air in the lower troposphere above the reference conditions (high \textbf{z$_{translation}$}), whereas the opposite is true for low \textbf{z$_{translation}$}. In this case, negative \textbf{dq/dt} in combination with negative diabatic forcing lead to a drying of moist conditions in the lower troposphere. Precipitation is insensitive to \textbf{z$_{translation}$} due to the small vertical extent of convective moistening, confined below 600 hPa; this latent node evidently avoids deep convective regimes. The generated temperature profiles of \textbf{z$_{translation}$} along latent dimension 3 are characteristic of the subtropics and mid-latitudes in the SP simulation. The \textbf{dT/dt} profiles show slight variations near the surface due to \textbf{z$_{translation}$}, while being insensitive in the middle troposphere. The fixed SST field (Figure S4) or the conditional averages of surface air temperatures (Figure S6) in certain regions can be used to gain a first visual orientation of the geographic origin of a generated sample. This first impression is complemented with a detailed search for such conditions in the SP test data. Furthermore, night-time conditions with small shortwave radiative heat flux at the model top \textbf{Q$_{sw\ top}$} and day-time conditions with high values of \textbf{Q$_{sw\ top}$} (\textbf{Q$_{sw\ top}$} $\sim$ 1000 $\frac{W}{m^2}$) can be distinguished for low \textbf{z$_{translation}$} and high \textbf{z$_{translation}$}, respectively.

Interestingly, the generated profiles and variables suggest that latent node 3 is mostly sorting information about sub-grid-scale processes \textbf{Y} within one sub-regime of \textbf{X}, rather than focusing on sorting the large-scale geographic variability in \textbf{X}. The strong response of \textbf{dq/dt} in the planetary boundary layer and adjacent layers, negligible precipitation formation and the characteristic temperature range between the subtropics and mid-latitudes, are key evidence that the latent node 3 encapsulates shallow convective processes. Shallow convection is influenced by the diurnal cycle, leading to a strengthening of shallow convective processes during the day and a weakening of these processes accompanied with a drying of the planetary boundary layer during the night, as it is supported by Figure \ref{fig:gen_Node_3}. 

\subsubsection{Latent Node 4: Mid Latitude Frontal Systems}

\begin{figure}[ht]
     \centering
     \includegraphics[width=13.5cm]{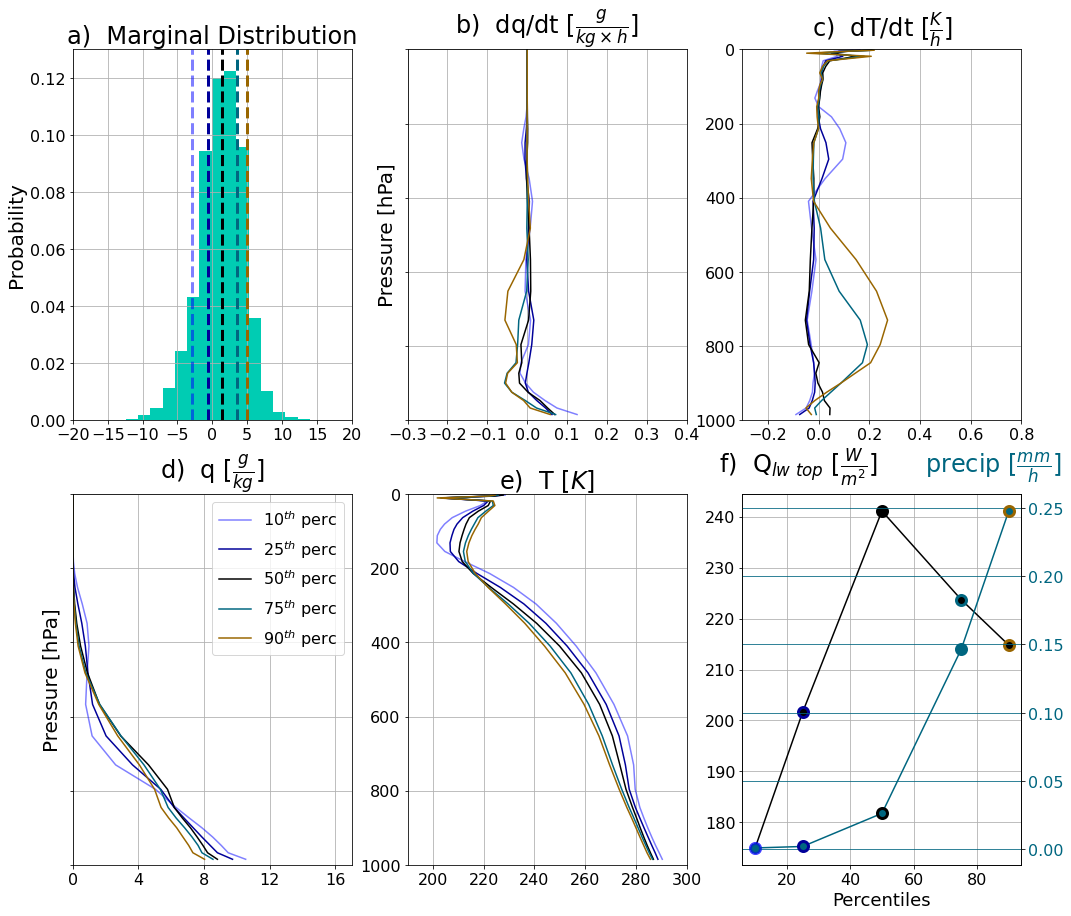}
     \caption{Marginal distribution of latent node 4 (a) and the resulting generated vertical profiles of specific humidity tendencies \textbf{dq/dt} (b), temperature tendencies \textbf{dT/dt} (c), specific humidity \textbf{q} (d) and temperatures \textbf{T} (e). The dashed lines in the marginal distribution plot represent the chosen percentiles (see legend in subplot d) and the resulting effect of the respective translation \textbf{z$_{translation}$} on the profiles is shown in the subplots. Furthermore, the longwave heat flux at the model top (\textbf{Q$_{lw \ top}$}) and the precipitation rate (\textbf{precip}) (f) are illustrated as function of the translation \textbf{z$_{translation}$} along the latent dimension 4. The marker-edge-color in panel f symbolise the respective percentiles of \textbf{z$_{translation}$}. The black lines in subplots b-e indicate the generated reference state with
     \textbf{z$_{median}$}.}
     \label{fig:gen_Node_4}
 \end{figure}

Mid-latitude frontal systems are characterized by a large variety of convective regimes associated with the warm or cold front of these systems \cite{Bony2015}. On latent node 4 we discover certain characteristic features in sub-grid-scale profiles \textbf{X} and associated large-scale fields \textbf{Y}. These features allow us to draw links to distinctive convective regimes of mid-latitude cyclones based on their fingerprint in \textbf{X} and \textbf{Y}. Unlike the previous latent nodes, the response of the latent node 4 (Node 4, Figure \ref{fig:gen_Node_4}) to the translation \textbf{z$_{translation}$} results in nearly constant solar insolation (\textbf{Q$_{sw \ top}$} $\sim$ 440 - 450 $\frac {W}{m^2}$, see Table S10) and a narrow meridional band. 

The generated surface temperature ranges from 286 K to 290 K with varying \textbf{z$_{translation}$}. This temperature range is common to mid-latitudes or the subtropics (e.g. see Figure S6) and can be found in the SPCAM simulations between 45$^{\circ}$ N / S and 25$^{\circ}$ N / S. Low \textbf{z$_{translation}$} corresponds to warmer and drier conditions in the free mid-troposphere between 800 hPa and 400 hPa, while moister conditions are found above and below. The anomalous moist conditions in the upper free troposphere are connected to a heating peak at 300 hPa (\textbf{dT/dt} $\sim$ 0.1 $\frac{K}{h}$, Figure \ref{fig:gen_Node_4}c). Likewise, the difference between the shortwave heat flux at the model top and the surface is relatively small (\textbf{Q$_{sw \ top}$ – Q$_{sw \ surf}$} $\sim$ 120 – 130 $\frac{W}{m^2}$, Table S10), which suggests optically thin clouds. Additionally, the outgoing long wave radiation is small (\textbf{Q$_{lw \ top}$} $<$ 200 $\frac{W}{m^2}$) and no precipitation is formed. These conditions are characteristic of high cirrus-like convection.

On the other side, high \textbf{z$_{translation}$} shows relatively strong heating tendencies in the free troposphere (\textbf{dT/dt} $>$ 0.2 $\frac{K}{h}$, see Figure \ref{fig:gen_Node_4}c) and drying conditions below 600 hPa down to the surface (\textbf{dq/dt} $\sim$ -0.1 $\frac {g}{kg \times h}$, Figure \ref{fig:gen_Node_4}b). These conditions, along with moderate precipitation (\textbf{precip} $\sim$ 0.15 - 0.25 $\frac {mm}{h}$), higher outgoing longwave heat flux (\textbf{Q$_{lw \ top}$} $>$ 200 $\frac{W}{m^2}$) and lower shortwave transmissivity (\textbf{Q$_{sw \ top}$ – Q$_{sw \ surf}$} $\sim$ 170 $\frac{W}{m^2}$, Table S10) characterize mid-level cumulus convection. 

Based on this evidence, we were able to show that latent node 4 focuses on sub-grid-scale convective processes in \textbf{Y}. The generated large-scale conditions exhibited by Node 4 are well-suited for these cirrus-like or cumulus convection regimes. In detail, latent node 4 shows a clear transition from a cirrus type convective regime (low \textbf{z$_{translation}$}) to a cumulus type precipitating convective regime ( high \textbf{z$_{translation}$}) in mid-latitudes. This response is associated with frontal systems, which consist of high cirrus clouds in the surroundings of the warm front and cumulus convection along the cold front \cite{Bony2015}.

\subsubsection{Latent Node 5: Deep Convection}

\begin{figure}[ht]
     \centering
     \includegraphics[width=13.5cm]{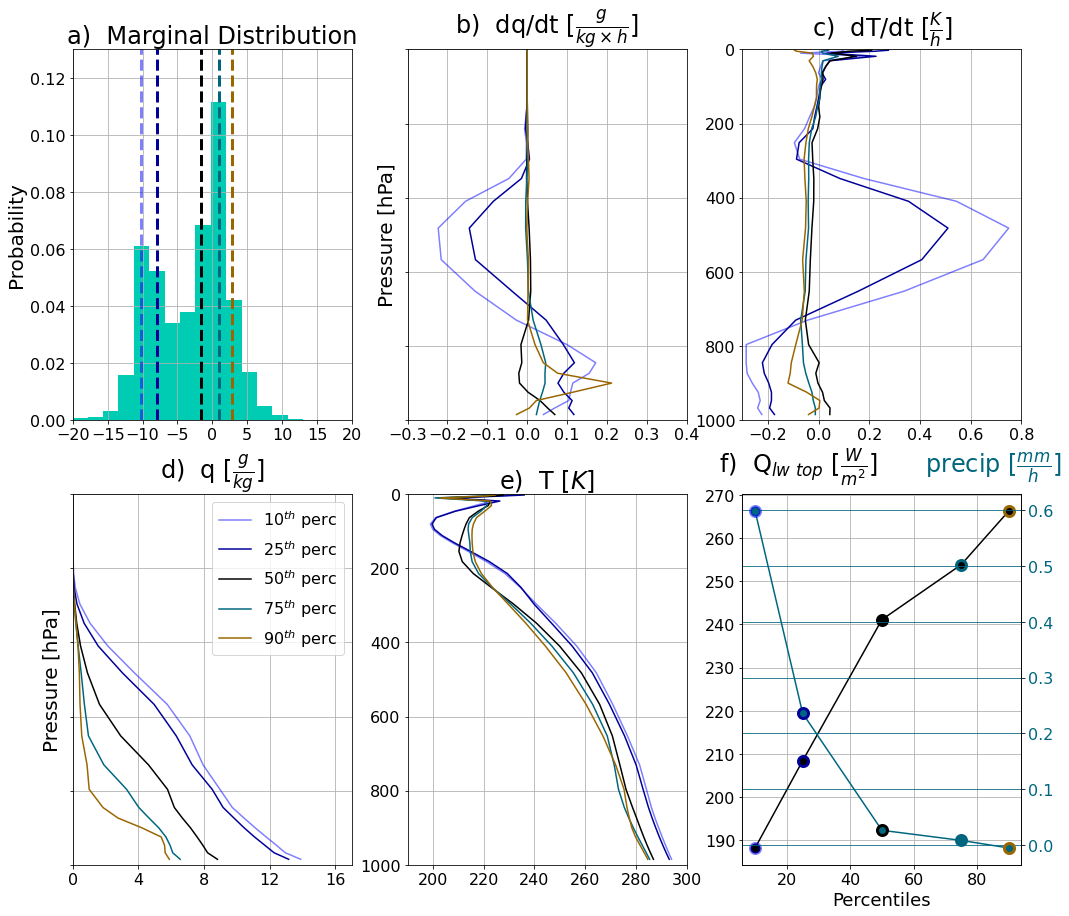}
     \caption{Marginal distribution of latent node 5 (a) and the resulting generated vertical profiles of specific humidity tendencies \textbf{dq/dt} (b), temperature tendencies \textbf{dT/dt} (c), specific humidity \textbf{q} (d) and temperatures \textbf{T} (e). The dashed lines in the marginal distribution plot represent the chosen percentiles (see legend in subplot d) and the resulting effect of the respective translation \textbf{z$_{translation}$} on the profiles is shown in the subplots. Furthermore, the longwave heat flux at the model top (\textbf{Q$_{lw \ top}$}) and the precipitation rate (\textbf{precip}) (f) are illustrated as function of the translation \textbf{z$_{translation}$} along the latent dimension 5. The marker-edge-color in panel f symbolise the respective percentiles of \textbf{z$_{translation}$}. The black lines in subplots b-e indicate the generated reference state with
     \textbf{z$_{median}$}.}
     \label{fig:gen_Node_5}
 \end{figure}

Deep convection is the cloud regime with the largest vertical extent. It is characterized by especially strong convective heating and drying throughout almost the entire troposphere, as can be seen in \citeA{Frenkel2015} and accompanied by anomalous intense precipitation (see Figure \ref{fig:VAE_clim_clim_conv_glob_multi}). The first mode of latent node 5 reveals general characteristics of a deep convective regime captured in generated sub-grid-scale variables \textbf{Y}. The response of latent dimension 5 (Node 5) to \textbf{z$_{translation}$} shows either strong deep convection (first mode in Figure \ref{fig:gen_Node_5}a) or stable conditions (second mode in Figure \ref{fig:gen_Node_5}a) in the troposphere. A surface temperature of 293 K for low \textbf{z$_{translation}$} indicates subtropical regions (e.g. the surface temperature in the tropics is at least 3 K warmer in this SPCAM simulation). The warmer and moister troposphere for low \textbf{z$_{translation}$} is accompanied with strong heating and drying tendencies peaking at around 500 hPa of 0.5 to 0.7 $\frac{K}{h}$ and -0.15 to -0.2 $\frac{g}{kg \times h}$ respectively. In this case, we observe intense precipitation formation up to 0.6 $\frac{mm}{h}$ and low outgoing longwave radiation (\textbf{Q$_{lw \ top}$} $<$ 201 $\frac{W}{m^2}$). All these conditions are characteristics of subtropical deep convective events. 

 In contrast, high \textbf{z$_{translation}$} is associated with a mid latitude surface air temperature (\textbf{T$_{surf}$} $\sim$ 5 K colder than in the subtropics). A night-time (Table S11), dryer troposphere with very small or negligible heating and moistening tendencies (manifestation of stable conditions) throughout the troposphere is accompanied with relatively large outgoing long wave radiation (\textbf{Q$_{lw \ top}$} $>$ 250 $\frac{W}{m^2}$) and no precipitation. Similar to latent node 3 and 4, latent node 5 comprises dominantly information about sub-grid-scale convective processes rather than large-scale geographic variability. Latent node 5 represents both deep convective events originating from the subtropics and mid-latitude stable conditions as can be already seen in the strong bimodality along the marginal distribution in Figure \ref{fig:gen_Node_5}a.

\section{Conclusion and Discussion}
 
\begin{figure}
    \centering
    \includegraphics[width=13.5cm]{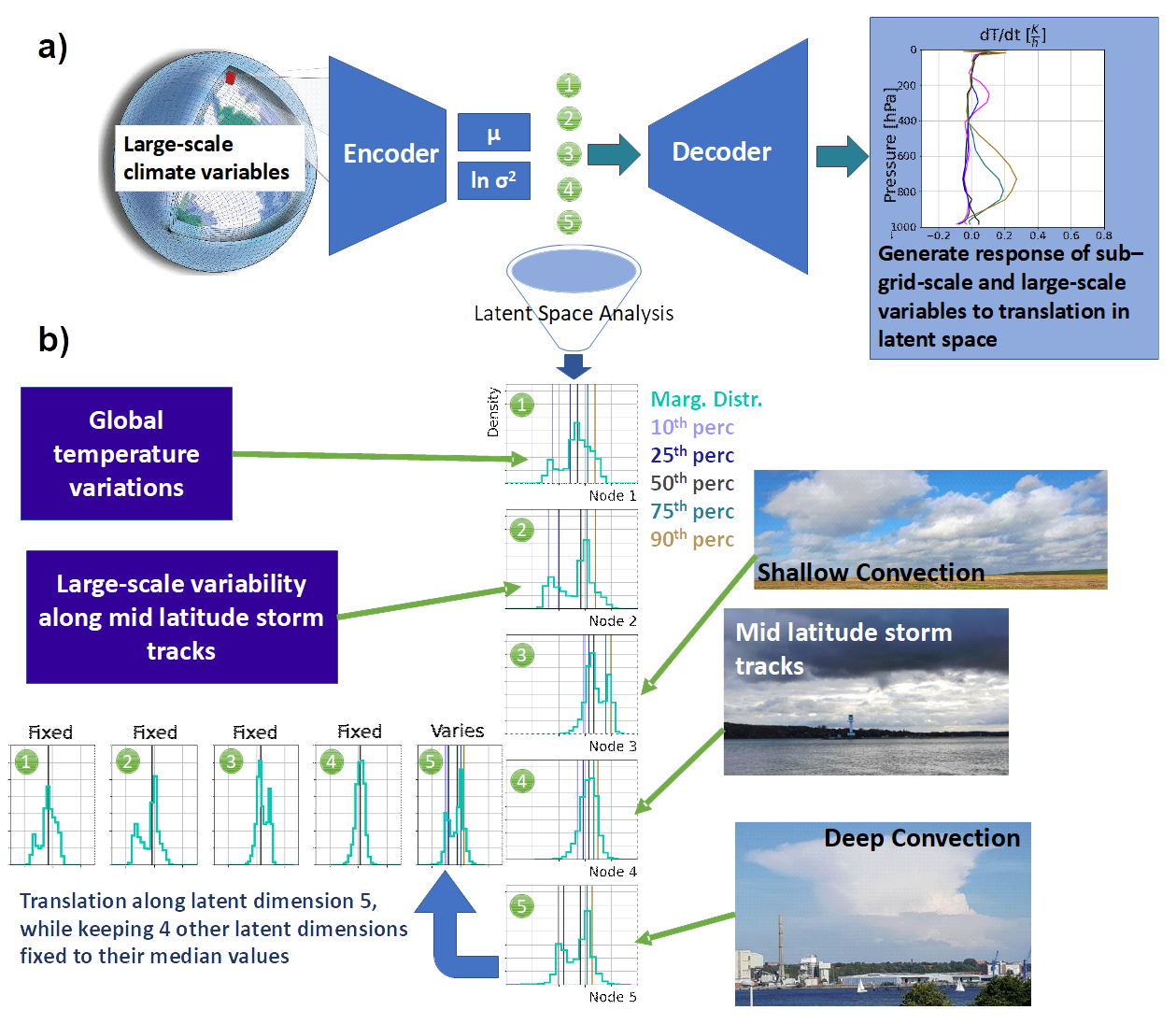}
    \caption{Schematic of the VED setup \textbf{(a)} the investigated convective regimes and drivers of convective processes in the latent space of VED for each node \textbf{(b)}. The translation along each latent dimension is shown in this example for Latent Node 5. The schematic of the large-scale atmospheric grid in \textbf{(a)} was adapted from \citeA{Schneider2017}.}
    \label{fig:VAE_summary_schematic}
\end{figure}

This study has shown how Variational Encoder Decoders (VEDs) can successfully machine learn a convective parameterization with considerable input compression while simultaneously enhancing the interpretability of deep learning methods, and enable better understanding of convective processes in climate models. We first showed that the VED is able to realistically reconstruct convective processes simulated by a superparameterized climate model, similar to previous studies with regular artificial neural nets (ANNs) \cite{Gentine2018,Rasp2018}, but using automatically compressed input data. Furthermore, we demonstrated that the VED also enhances the interpretability of the relationship between large-scale climate fields and sub-grid-scale convective variables via its latent manifold, which is unfeasible via ANNs without attribution methods due to ANNs' large dimensionality (large number of hidden layers and nodes per layer). Our analysis is based on 9 months (equally split into training, validation and test data) of an aquaplanet simulation of the superparameterized community atmosphere model (SPCAM). As shown in Figure \ref{fig:VAE_summary_schematic}a, the input variables of the VED resembled the large-scale climate fields (temperature, specific humidity and other thermodynamic drivers) from the general circulation model (CAM) passed onto the embedded cloud resolving model (SP). The latent space (lower dimensional manifold inside the network) of the VED had a dimensionality of five nodes, which is a small fraction of the dimensionality of the original input nodes information. To create an interpretable latent space, our optimal network reconstructed a combination of sub-grid-scale convective variables related to the SP component and large-scale climate variables associated with CAM. In comparison, as we have shown in the supplemental material, VEDs that attempt the traditional mapping from \textbf{X} to \textbf{Y} alone turn out to be less amenable to latent space exploration.

As a first step, we evaluated the reproduction performance of convective processes of the VED against a reference ANN \cite{Rasp2018}. The VED was capable of reconstructing the mean statistics of sub-grid-scale convective variables with an overall comparable, though slightly decreased, skill than the reference ANN despite the strong dimensionality reduction down to five latent nodes. This speaks to the dimensionality of information content required for a convective parameterization, and associated trade-offs. We found that compressing the input information did not overly distort the tropical wave spectrum. We showed that the choice of the latent space width is a critical hyperparameter for reproduction skills. Larger latent space widths ($\sim$ 8 nodes) yielded a reproduction performance of convective processes with almost the skill of the reference ANN, while smaller latent space widths ($\sim$ 2 nodes) still enabled an improved reproduction compared to a multi-dimensional linear regression baseline. We chose a latent space of five nodes as a sensible compromise between reproduction abilities of convective regimes and sensitivities separable in the latent manifold.

We began the analysis towards our main interest - latent space exploration with respect to physical interpretability – using traditional methods visualizing physical properties in a 2D projection of its leading PCs. This revealed that the VED distinguished day- and night-time conditions and varying strength of convective processes using the precipitation rate and outgoing longwave radiation as a proxy in its latent space (which was 2D compressed with a PCA for the purpose of visualization). The VED separated different global climate conditions and associated convective regimes from the poles to the equator in its latent space. The realistic reproduction of convective processes and climate conditions, along with the encapsulated information on geographic variability in an interpretable latent manifold, allowed a detailed analysis of governing drivers of convection and convective regimes with a VED.

Our latent exploration was then deepened by investigating convective processes and related drivers via a generative modeling approach, i.e., forcing the decoder with the variability encapsulated along each latent dimension. The resulting temperature, specific humidity, heating, and moistening profiles successfully separated  well-known large-scale driving climate conditions and convective regimes. Figure \ref{fig:VAE_summary_schematic}b summarizes the main results of this generative modeling approach. Overall, convective processes are controlled by large meridional gradients in temperature and specific humidity, from the equator to the poles, which were captured by the VED's Node 1 (Figure \ref{fig:VAE_summary_schematic}). We identified the large-scale climate variability in specific humidity and temperatures along the mid-latitude storm tracks (Node 2, Figure \ref{fig:VAE_summary_schematic}) as the other major driver of convective processes. Daytime stable, cold and dry tropospheric conditions suppress convective processes in the entire troposphere, whereas night-time unstable, warm and moist conditions in the troposphere drive precipitating mid-level convection. Apart from these large-scale nodes, the VED further reveals characteristics of distinct convective regimes on the remaining 3 latent nodes. The VED confined shallow convective processes below 600 hPa within its Node 3 (Figure \ref{fig:VAE_summary_schematic}); these processes are generally driven by surface diabatic fluxes and are predominantly originating from mid-latitudes and the subtropics. In anomalous dry conditions, positive surface diabatic heat fluxes during day-time enhance shallow convective processes associated with a convective moistening of the lower troposphere. The opposite is true in anomalous wet conditions during night-time.  The mid-latitude storm tracks show large variability with respect to convective regimes associated with the eastward migrating frontal systems, features that were captured in the VED’s latent space (Node 4, Figure \ref{fig:VAE_summary_schematic}). In the surroundings of the warm front high, optically thin, non-precipitating cirrus-like convection is found. In contrast lower, optically thick cumulus-like convection with intermediate precipitation formation is predominant near the cold front. Furthermore, deep convective regimes in the subtropics were clearly captured by the VED (Node 5, Figure \ref{fig:VAE_summary_schematic}). In this case, convective processes extend in the entire troposphere with a pronounced convective heating and drying near 500 hPa and are associated with intense precipitation. Opposing this extreme convective case, we found night-time, stable, cold and dry conditions in the free troposphere, which suppress convective processes on the other side of Node 5. Finally, while the interpretation of these convective regimes always required domain knowledge, the generative modeling approach simplified the analysis in comparison to other statistical analysis tools (e.g., correlations, clustering, attribution methods).

Repeating this analysis with an Encoder Decoder (ED) yielded almost identical reproduction capabilities compared to the VED, but the ED's latent space was significantly harder to interpret, with less pronounced center of actions for a given variable (see Figure S5 and S7). This hindered the identification of convective regimes or large-scale drivers of convective predictability within the latent space of ED. For example, although the ED captured a cirrus-like regime, no cumulus or deep convective regimes could be found with the generative modeling method. Likewise, the connection between large-scale climate variables was often less pronounced for the ED, which resulted in larger uncertainties of the geographic origin of a specific sample compared to the VED.

We discovered convective regimes with the VED that are in general agreement with existing work focused on tropical convection \cite{Frenkel2012, Frenkel2013,Frenkel2015,Huaman2017}. The specific humidity profile of the shallow convective regime of the VED was largely similar to the observed shallow convective latent heating profile in \citeA{Huaman2017} with a heating peak around 800 hPa. Furthermore, the heating profile of the mid-latitude cirrus-like regime of the VED compared well with that of the tropical stratiform regime shown in \citeA{Frenkel2015} despite strong differences in the ambient conditions that led to their formation. Also the heating profiles of the mid-latitude cumulus regime of the VED and their tropical congestus expressed similarities in the lower troposphere with a pronounced convective heating peak above the boundary layer. Likewise, the VED’s subtropical and tropical deep convection regime of \citeA{Frenkel2015} were characterized by similar heating profiles. In our case, we identified these regimes solely based on SPCAM data in the latent space of the VED, where we did not prescribe the characteristics of each convective regime like it was done in the multi-cloud approach presented in \citeA{Frenkel2012} and adapted from \citeA{Khouider2006}. Furthermore, our approach was not based on inferred heating profiles via subclassing precipitation regimes (Stratiform, Convective, Shallow) as it was done for observational satellite products in \citeA{Huaman2017}. 

This work presented how convective processes, convective regimes, and large-scale drivers of convection in climate models can be investigated by leveraging generative machine learning (ML) approaches. Our approach enhanced the understanding of acting convective processes and the corresponding large-scale environment in which they form. As a next step, one could study cirrus-like or cumulus convection in detail by, for example, separating specific humidity and moistening tendencies related to the ice phase, linking how microphysical processes influence convection and are, in turn, affected by climate conditions (i.e., formation of ice phase, mixed phase or liquid phase clouds). Likewise, the development of regime-oriented ML-based convection parameterizations appears to be achievable with generative deep learning methods. Finally, VEDs could play an essential role in constructing new stochastic convection parameterizations, which could improve the representation of clouds and convection in Earth System Models. Our results suggest that VED representations of climate processes can effectively combine statistical prediction with data-driven analysis, paving the way towards machine learning-based Earth System Models that remain interpretable, albeit through the yet mostly unfamiliar eccentricities of latent space exploration.

\section{Open Research}

The code used to train the VEDs, the conditional VAE and reference models, and to produce all figures of this manuscript is accessible in the following Github repository: \url{https://github.com/EyringMLClimateGroup/behrens22james_SPCAM_VED}, which is archived with Zenodo (\url{https://zenodo.org/record/6882230#.YtqJr4TP25c}). The repository includes the Jupyter Notebooks, python files, conda environments used to reproduce all figures of the manuscript and attached supporting information. The text file \url{https://github.com/EyringMLClimateGroup/behrens22james_SPCAM_VED/blob/master/List_of_Figures.txt} illustrates where to find the code to reproduce each Figure in the Github repository. 
The above mentioned Github repository is a fork of Stephan Rasp's main repository published for \citeA{Rasp2018}, which can be found here: \url{https://github.com/raspstephan/CBRAIN-CAM}, archived using Zenodo (\url{https://zenodo.org/record/1402384#.YajSg9BKiUk}). The repository includes a helpful quickstart guide \url{https://github.com/raspstephan/CBRAIN-CAM/blob/master/quickstart.ipynb} to preprocess raw SPCAM data, train a neural network similar to reference ANN and how to evaluate it.

An example of SPCAM data was archived on Zenodo for \citeA{Rasp2018} and can be found here: \url{https://zenodo.org/record/2559313#.YlVG0tPP25c} . The full SPCAM raw data, of the order of several TBs, is archived on the GreenPlanet cluster at UC Irvine and available upon request. Additionally the preprocessed SPCAM data, of the order of 1 TB, used in this study is archived on DKRZ and is also available upon request.

%
%
%
%
%
%
%
%

\acknowledgments

We thank two anonymous reviewers and one internal reviewer for their helpful comments and suggestions, which improved our manuscript.
Funding for this study was provided by the European Research Council (ERC) Synergy Grant “Understanding and modeling the Earth System with Machine Learning (USMILE)” under the Horizon 2020 research and innovation programme (Grant agreement No. 855187). Beucler acknowledges funding from the Columbia University sub-award 1 (PG010560‐01). Gentine acknowledges funding from the National Science Foundation Science and Technology Center, Learning the Earth with Artificial intelligence and Physics, LEAP. Pritchard acknowledges funding from the National Science Foundation Science and Technology Center, Learning the Earth with Artificial intelligence and Physics (LEAP) and from the US Department of Energy Advanced Scientific Computing Research program (DE-SC0022331). This work used resources of the Deutsches Klimarechenzentrum (DKRZ) granted by its Scientific Steering Committee (WLA) under project ID 1179 (USMILE).


%
%

\bibliography{main_behrens_22.bbl}

%
%
%
%
%

\end{document}


%
%


\title{Supporting Information for ``Non-Linear Dimensionality Reduction with a Variational Encoder Decoder to Understand Convective Processes in Climate Models''}
%
%

%
%



\authors{Gunnar Behrens \affil{1,2}, Tom Beucler \affil{3}, Pierre Gentine \affil{2,4}, Fernando Iglesias-Suarez \affil{1}, Michael Pritchard \affil{5}, Veronika Eyring \affil{1,6}}

\affiliation{1}{Deutsches Zentrum für Luft- und Raumfahrt (DLR), Institut für Physik der Atmosphäre, Oberpfaffenhofen, Germany}
\affiliation{2}{Department of Earth and Environmental Engineering, Columbia University, New York, NY 10027, USA}
\affiliation{3}{Institute of Earth Surface Dynamics, University of Lausanne, Lausanne, VD 1015, Switzerland}
\affiliation{4}{Earth Institute and Data Science Institute, Columbia University, New York, NY 10027, USA}
\affiliation{5}{Department of Earth System Science, University of California Irvine, Irvine, CA, USA}
\affiliation{6}{University of Bremen, Institute of Environmental Physics (IUP), Bremen, Germany}

%
%

%

\begin{article}

%
%

\noindent\textbf{Contents of this file}
\begin{enumerate}
\item Tables S1 to S11
\item Figures S1 to S19

\end{enumerate}

\noindent\textbf{Introduction}

The supporting information are structured as follows and each section can be read individually: 

In section S.1 we show the hyperparameters of VED and explain how we conducted the search for a suitable set of hyperparameters of VED. Furthermore we discuss the used VED output normalization dictionary. In section S.2 we show additional figures for the general evaluation of VED and other reference networks. This section further describes differences in reproduction skill if either the VED or the output normalization of \citeA{Rasp2018} is used. Furthermore we describe differences in the interpretability between the VED's latent space and a principal component analysis on the large-scale variables in this section. Also we show that the latent space exploration with conditional averages can be conducted on the five original latent dimension. Section S.3 shows one alternative VED and a conditional VAE structure and discusses their strengths and limitations. We describe in subsection A) the VED$_{X \rightarrow Y}$ and in subsection B) a conditional VAE (cVAE). Section S.4 comprises the tables of all generated 2D SP or CAM variables with our generative modeling approach. Additionally the squared Pearson correlation coefficients R² between the latent nodes and vertical heating, moistening, specific humidity and temperature profiles in space-time and time are shown in this section respectively.

\clearpage

\noindent\textbf{S.1 VED Hyperparameters based on a Hyperparameter Search and Normalisation}

\begin{table}
    \centering
    \begin{tabular}{p{3.8cm}|p{8cm}}
         \textbf{Hyperparameter of VED} & \textbf{Values}\\
         \hline
         \hline
         Learning Rate & 0.00074594 \\
         \hline
         Training / learning rate decrease & 40 epochs, learning decrease every 7$^{th}$ epoch by factor 5 \\
         \hline
         Batch size	& 714\\
         \hline
        Latent Space Width & 5 nodes \\
        \hline
        Node Size of Encoder &	[64,463,463,232,116,58,29,5]\\
        \hline
        Node Size of Decoder &	[5,29,58,116,232,463,463,129]\\
        \hline 
        Activation Encoder &
        [Input, ReLU, ReLU, ReLU, ReLU, ReLU, ReLU, Lambda]\\
        \hline
        Activation Decoder &
        [Input, ReLU, ReLU, ReLU, ReLU, ReLU, ReLU, ELU]\\
        \hline
        KL Annealing &	Linear annealing from 2$^{nd}$ to 7$^{th}$ epoch 
    \end{tabular}
    \caption{Hyperparameters and architecture of the final VED which uses large-scale CAM variables \textbf{X} to investigate simulated convective processes of SP \textbf{Y} together with driving climate conditions.}
    \label{tab:VAE_final_hyper}
\end{table}

\begin{table}
    \centering
    \begin{tabular}{p{8cm}|p{4cm}}
        \textbf{Hyperparameter range of VAE$_{X \rightarrow X}$} & \textbf{Values}\\
        \hline
        \hline
        Initial learning rate & 	10$^{-5}$ to 5 $\times$ 10$^{-4}$ \\
        \hline
        Batch size & 200 to 8192 \\
        \hline
        Latent Space Width & 2 to 5 nodes\\
        \hline
        Node Size of first or last hidden layer of Encoder or Decoder &	300 to 500 \\
        \hline 
        Depth of Encoder or Decoder in hidden layers  &  5 to 7 hidden layers 
    \end{tabular}
    \caption{Hyperparameter range of search for initial VAE, which reproduces large-scale climate variables \textbf{X} with \textbf{X} as input data set. The hyperparameter search was conducted over 120 trials and 30 epochs with a learning rate decrease after every 5$^{th}$ epoch by a factor 5.}
    \label{tab:VAE_X_X_hyper}
\end{table}

\begin{table}
    \centering
    \begin{tabular}{p{5cm}|p{6cm}}
        \textbf{Hyperparameter range of VED} & \textbf{Values}\\
        \hline
        \hline
        Initial learning rate & 	5 $\times$ 10$^{-5}$ to 5 $\times$ 10$^{-3}$ \\
        \hline
        Batch size & 200 to 8000 
    \end{tabular}
    \caption{Hyperparameter range of search for VED, the main model in this study. The hyperparameter search was conducted over 80 trials and 20 epochs with one learning rate decrease after the 10$^{th}$ epoch.}
    \label{tab:VAE_hyper}
\end{table}

In earlier experiments we found that the output normalisation used in \citeA{Rasp2018} was not well-suited for the optimization of a VED during training. With their output normalisation dictionary, the VED focused solely on the reproduction of radiative fluxes in \textbf{Y} and lacked skill with respect to heating and moistening profiles. Therefore we had to re-scale the output normalisation dictionary for \textbf{Y} and implement a suitable scaling for the extended variable list \textbf{O}. The vertical profiles of temperature, specific humidity and specific humidity tendency are normalised by long-term (3 month) standard deviations of the near surface model level. In the case of temperature tendencies, the standard deviation on the 845 hPa level is used due to the dominant variability of convection related temporal temperature changes on this level near the upper limit of the planetary boundary layer in SP data. The remaining 2D variables of radiative properties, precipitation rates and surface pressure are standardised.

We initially performed a hyperparameter search (random search) with 120 trials for a VAE$_{X \rightarrow X}$,which was trained on large-scale climate variables \textbf{X} to reproduce X. Table \ref{tab:VAE_X_X_hyper} shows the hyperparameter range for a hyperparameter search over a sequence of 1 month of SP data.

The best-performing encoder and decoder hyperparameter settings; 6 hidden layers, 463 nodes in the first and last hidden layer and a latent space width of 5 nodes; were fixed for the development of the VED presented in the paper. To account for shifts in suitable learning rates and batch size due to the additional sub-grid-scale output variables \textbf{Y}, we conducted a second hyperparameter search (random search) for our main VED specifically over a sequence of 1 month of SP data, see Table \ref{tab:VAE_hyper}.

After that, we fixed the initial learning rate and batch size and conducted further sensitivity tests with respect to the latent space width (which are documented in section 3) of VED. The choice of activation functions in the hidden layers is based on small initial experiments with VED, which showed enhanced emulation skill if the last hidden layer was elu-activated (exponential linear unit). 

\noindent\textbf{S.2 Evaluation of VED and the Reference Networks}

\begin{table}
    \centering
    \begin{tabular}{c|c|c|c}
        \textbf{Network} & \textbf{Training MSE} & \textbf{Validation MSE} & \textbf{Test MSE}\\  
        \hline
        \hline
        VED	& 0.162	& 0.165	& 0.165\\
        \hline
        EDD	& 0.162	& 0.165 &	0.165\\
        \hline
        LR	& 0.242	& 0.244 &	0.243\\
        \hline
        ANN	& 0.133	& 0.135 &	0.135\\
    \end{tabular}
    \caption{Mean squared errors (MSE) of predicted sub-grid-scale SP variables \textbf{Y} of the VED, ED, LR, reference ANN on the training, validation and test data sets (3 month of SP data) using the VED output normalization.}
    \label{tab:Mse_Results}
\end{table}

\begin{figure}
    \centering
    \includegraphics[width=12cm]{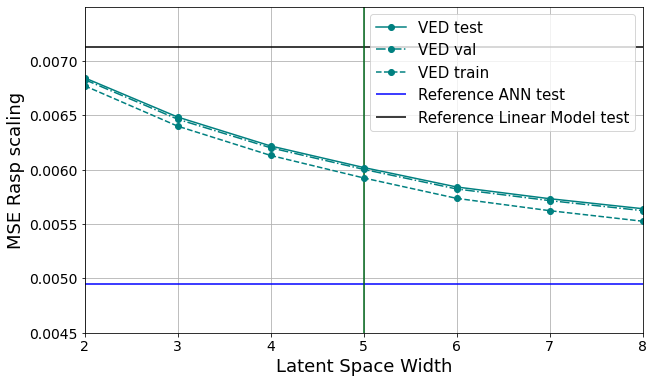}
    \caption{Similar to Figure 2, mean squared error as a function of Latent Space Width of the VED for the test (solid cyan), validation (dashed-dotted cyan) and training data set (dashed cyan curve) using the output normalization of the reference ANN \cite{Rasp2018} as y-axis. The horizontal solid blue / black line represents the MSE scores of the reference ANN \cite{Rasp2018} / a linear version of this network (Reference Linear Model) on test data with fixed layer width of 256 nodes in the 9 hidden layers.}
    \label{fig:VAE_BN_Rasp}
\end{figure}

If we use the output normalization of reference ANN to investigate the sensitivity of the VED performance as a function of latent space width, then we observe similar asymptotic behaviour as in Figure 2, see Figure \ref{fig:VAE_BN_Rasp}. The VED shows an improved emulation skill compared to the reference linear model with fixed layer widths of 256 nodes. The difference in performance between the VED and reference ANN increases if the output normalization of \citeA{Rasp2018} is used, which points to the fact that the VED output scaling weights SP variables \textbf{Y} differently. The VED has a decreased performance compared to reference ANN, but is converging to a similar level of emulation with increasing latent space width.

\begin{figure}
    \centering
    \includegraphics[width=13.5cm]{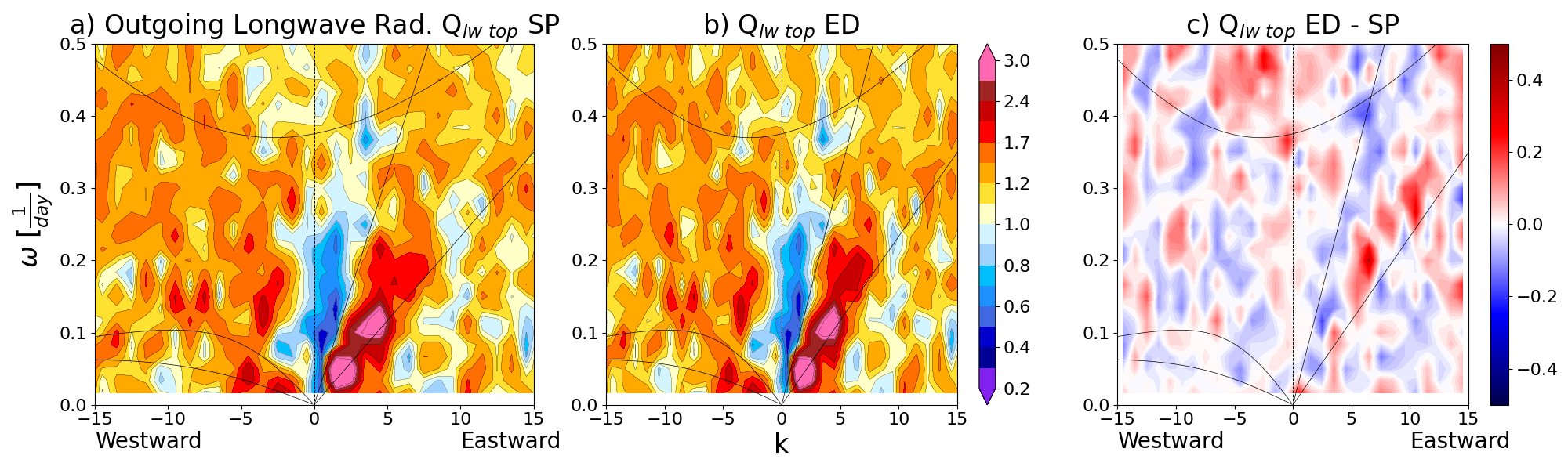}
    \caption{Wheeler Kiladis diagram based on tropical outgoing longwave radiation [15$^{\circ}$ N - 15$^{\circ}$ S] of SP (a), of ED predictions (b) and the absolute difference of spatio-temporal wave spectra ED - SP (c) for 1 year of SP data.}
    \label{fig:W_K_AED}
\end{figure}

\begin{figure}
    \centering
    \includegraphics[width=13.5cm]{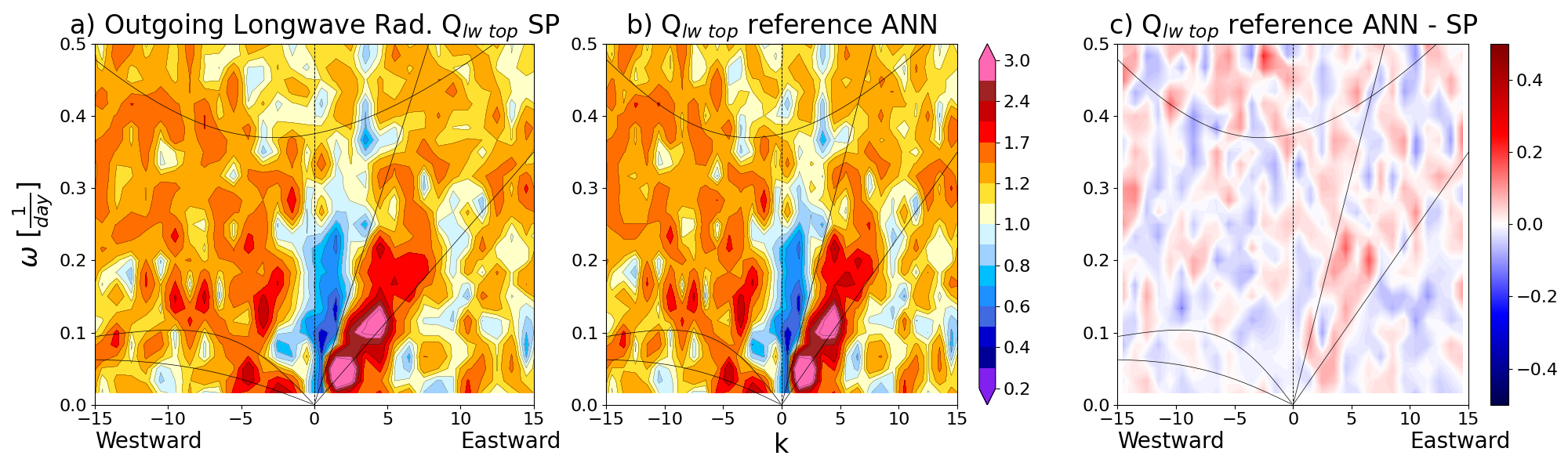}
    \caption{Wheeler Kiladis diagram based on tropical outgoing longwave radiation [15$^{\circ}$ N - 15$^{\circ}$ S] of SP (a), of reference ANN predictions (b) and the absolute difference of spatio-temporal wave spectra reference ANN - SP (c) for 1 year of SP data.}
    \label{fig:W_K_ref_ANN}
\end{figure}

\begin{figure}
    \centering
    \includegraphics[width=9.5cm]{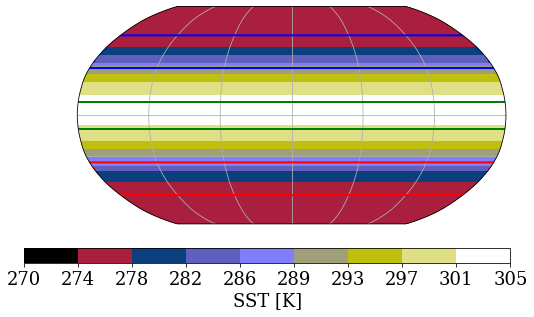}
    \caption{Fixed Sea Surface Temperature (SST) forcing of the SPCAM simulation following \citeA{Andersen2012}. The blue / red zonal lines indicate the region of Northern / Southern mid latitudes between 60$^{\circ}$ N/S and 35$^{\circ}$ N/S. The green lines indicate the deep tropics with the ITCZ between 10$^{\circ}$ S and 10$^{\circ}$ N.}
    \label{fig:SST}
\end{figure}

\begin{figure}
    \centering
    \includegraphics[width=6.5cm]{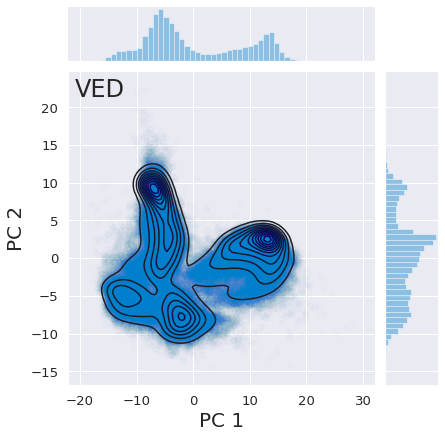}
    \includegraphics[width=6.5cm]{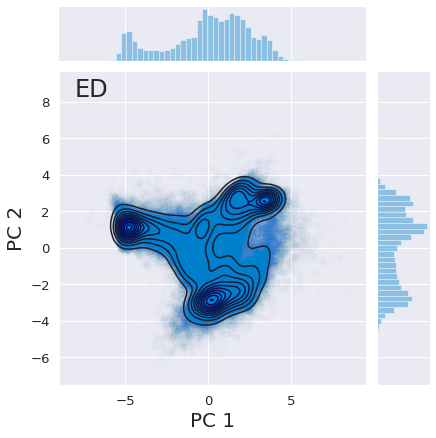}
    \caption{Scatter plot with isolines and histograms of the Joint and Conditional distributions of the PCA compressed latent space of VED (left) and ED (right). The plot is based on 100000 randomly picked samples from CAM test data. The 1$^{st}$ / 2$^{nd}$ PC of the resulting compressed latent space is the x-axis / y-axis in the respective subplot.}
    \label{fig:VAE_AED}
\end{figure}

\begin{figure}
    \centering
    \includegraphics[width=11.5cm]{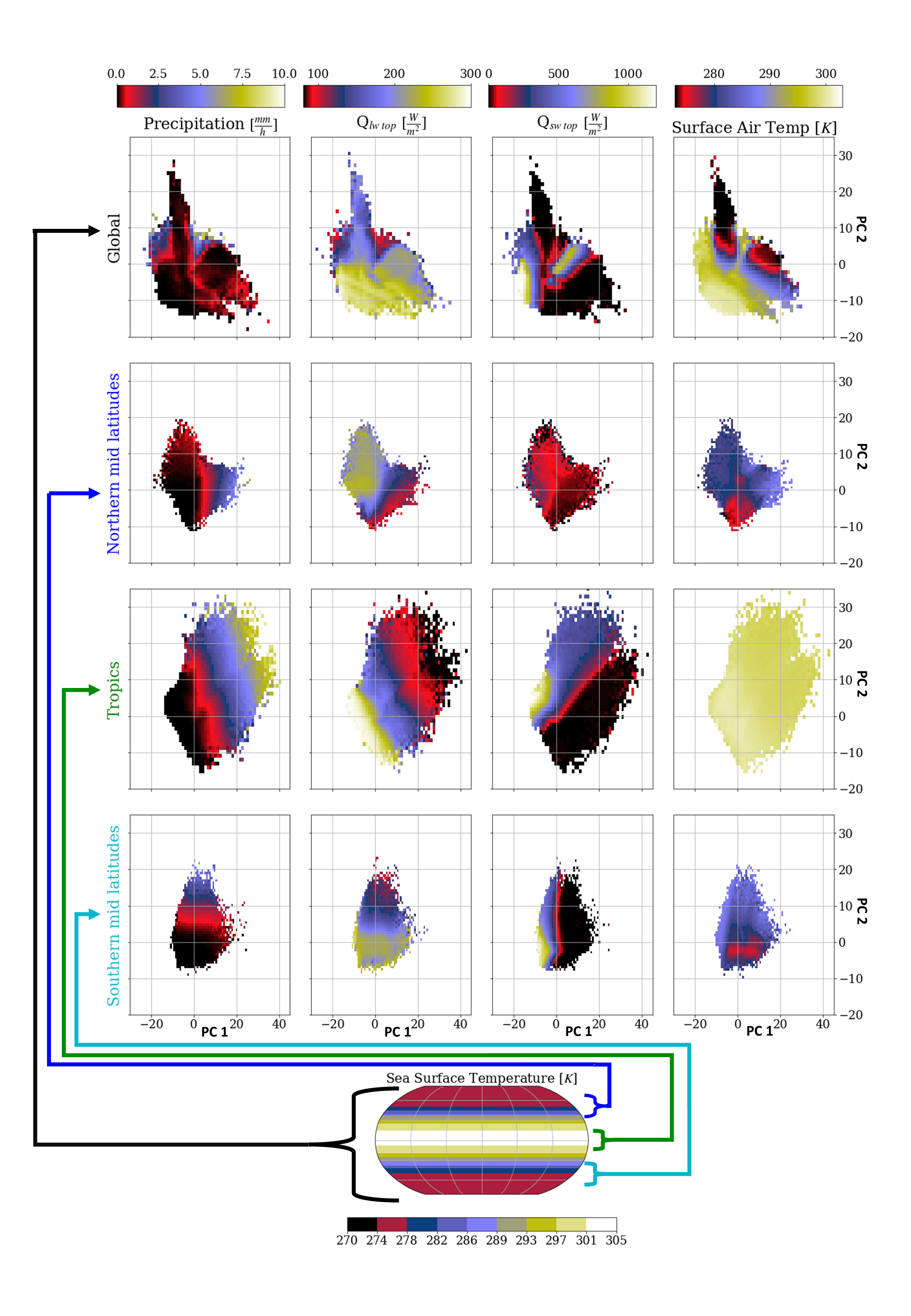}
    \caption{Latent Space clustering of VED for precipitation (left), outgoing longwave radiation (\textbf{Q$_{lw \ top}$}) (left middle), shortwave heat flux at the model top (\textbf{Q$_{sw \ top}$}) (right middle) and Surface Air Temperature (\textbf{T$_{surf}$}) (right column). The first row illustrates the clustering in the PCA compressed latent space with respect to the SP / CAM variables on global scales (as seen in Figure 4). The lower rows depict the Latent Space clustering in the evaluated regions Northern Mid Latitudes (2$^{nd}$ row), Tropics (3$^{rd}$ row) and Southern Mid Latitudes (4$^{th}$ row). The x-axis / y-axis represents the 1$^{st}$ / 2$^{nd}$ leading PC of the global / regional latent space.}
    \label{fig:VAE_port}
\end{figure}

\begin{figure}
    \centering
    \includegraphics[width=11.5cm]{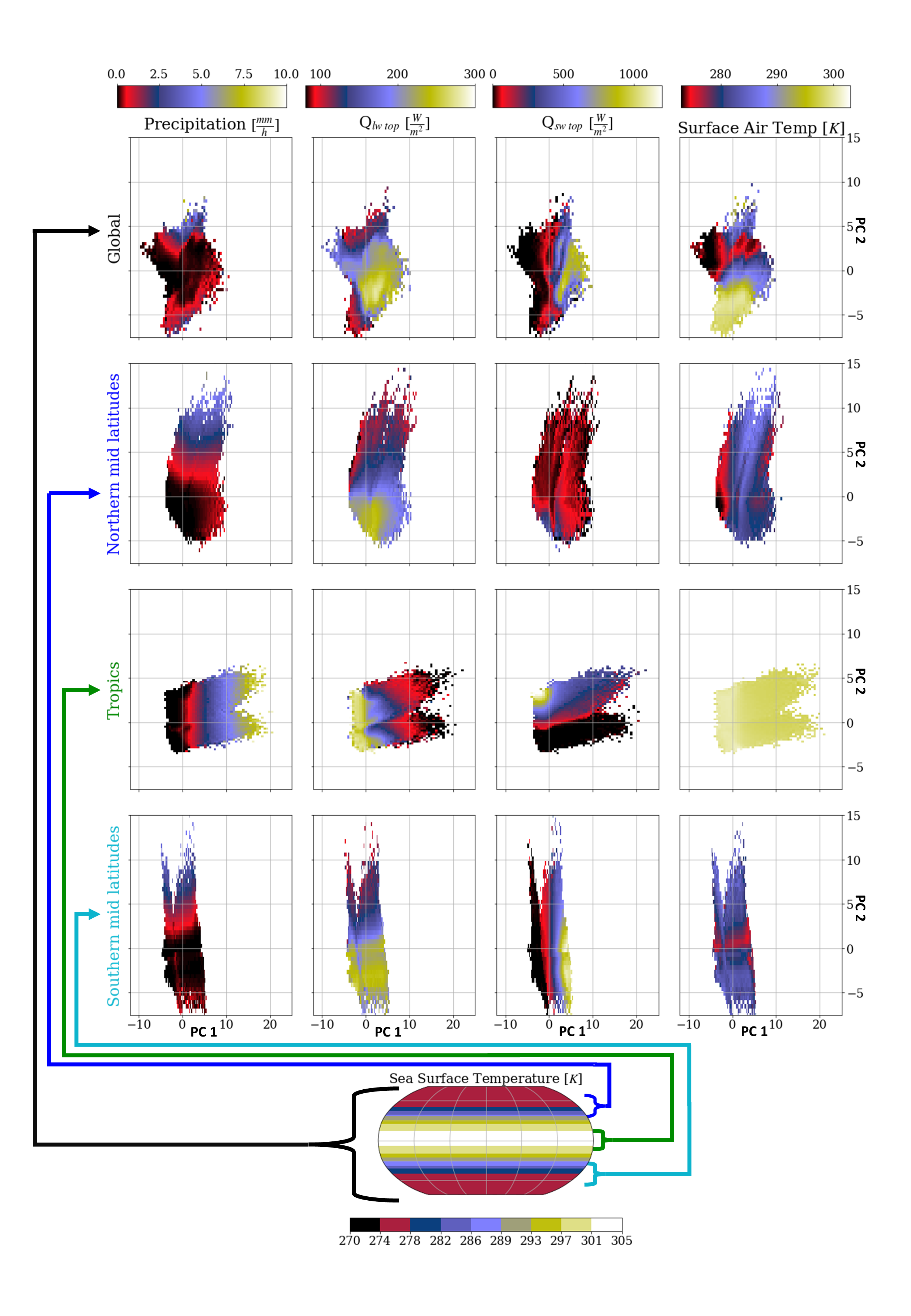}
    \caption{Latent Space clustering of ED for precipitation (left), outgoing longwave radiation (\textbf{Q$_{lw \ top}$}) (left middle), shortwave heat flux at the model top (\textbf{Q$_{sw \ top}$}) (right middle) and Surface Air Temperature (\textbf{T$_{surf}$}) (right column). The first row illustrates the clustering in the PCA compressed latent space with respect to the SP / CAM variables on global scales (as seen in Figure 4). The lower rows depict the Latent Space clustering in the evaluated regions Northern Mid Latitudes (2$^{nd}$ row), Tropics (3$^{rd}$ row) and Southern Mid Latitudes (4$^{th}$ row). The x-axis / y-axis represents the 1$^{st}$/ 2$^{nd}$ leading PC of the global / regional latent space.}
    \label{fig:AED_port}
\end{figure}

\begin{figure}
    \centering
    \includegraphics[width=15cm]{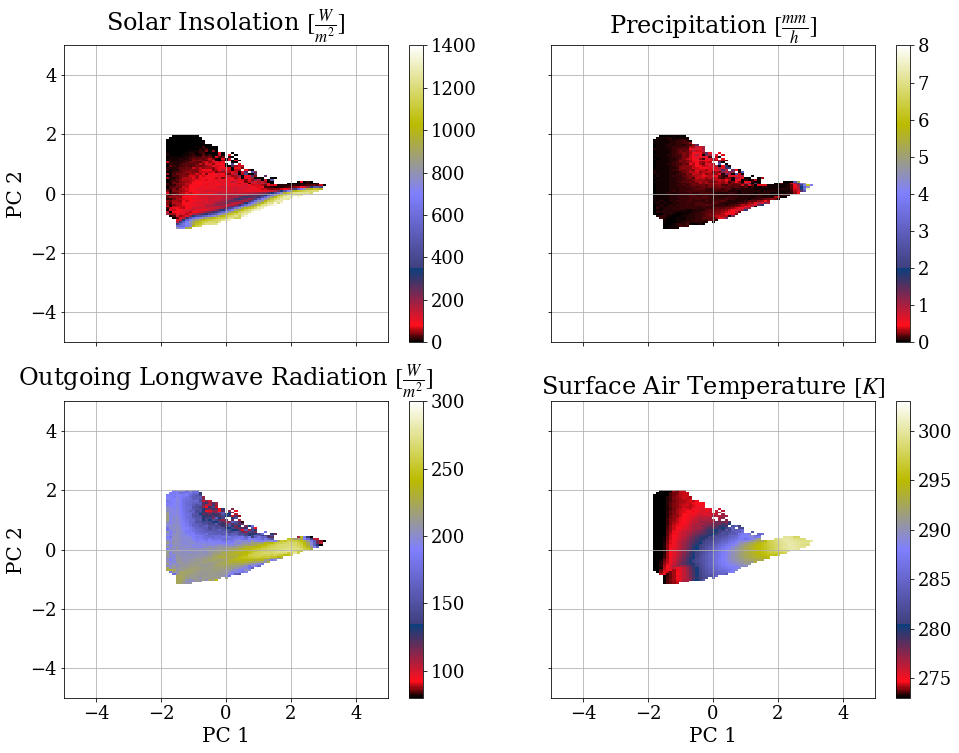}
    \caption{Conditional averages of solar insolation (upper left), precipitation (upper right), outgoing longwave radiation (lower left) and surface air temperature (lower right panel) in the submanifold spanned by the first two leading PC's of the large-scale variables \textbf{X}. Similar to Figure 5 the conditional averages are computed based on 1000000 randomly selected samples from the test data set. }
    \label{fig:my_label}
\end{figure}

\begin{figure}
    \centering
    \includegraphics[width=12cm]{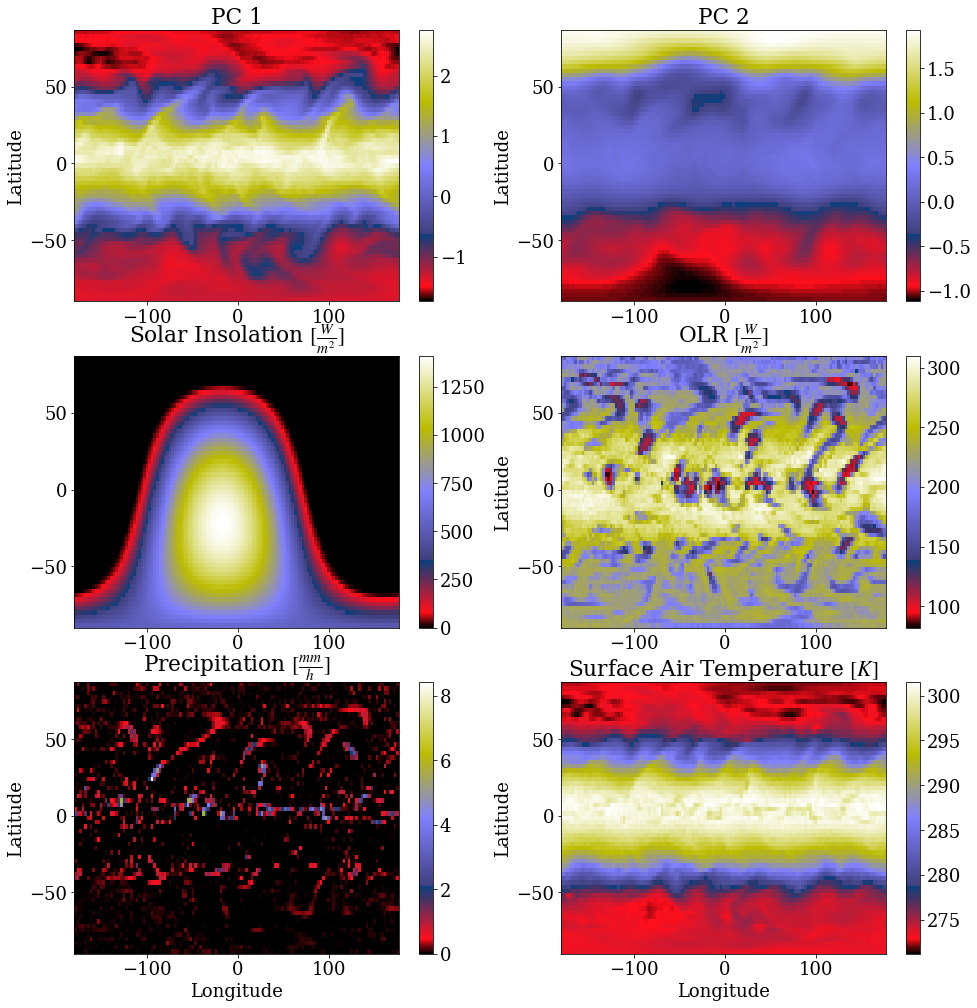}
    \caption{Latitude-Longitude plot of the first (upper left, PC1) and second (upper right, PC2) leading principal component of a PCA on the large-scale variables \textbf{X} and respective large-scale and sub-grid-scale variables of the test data set for a particular time step.}
    \label{fig:PCA_input_snap_shot}
\end{figure}

\begin{figure}
    \centering
    \includegraphics[width=15cm]{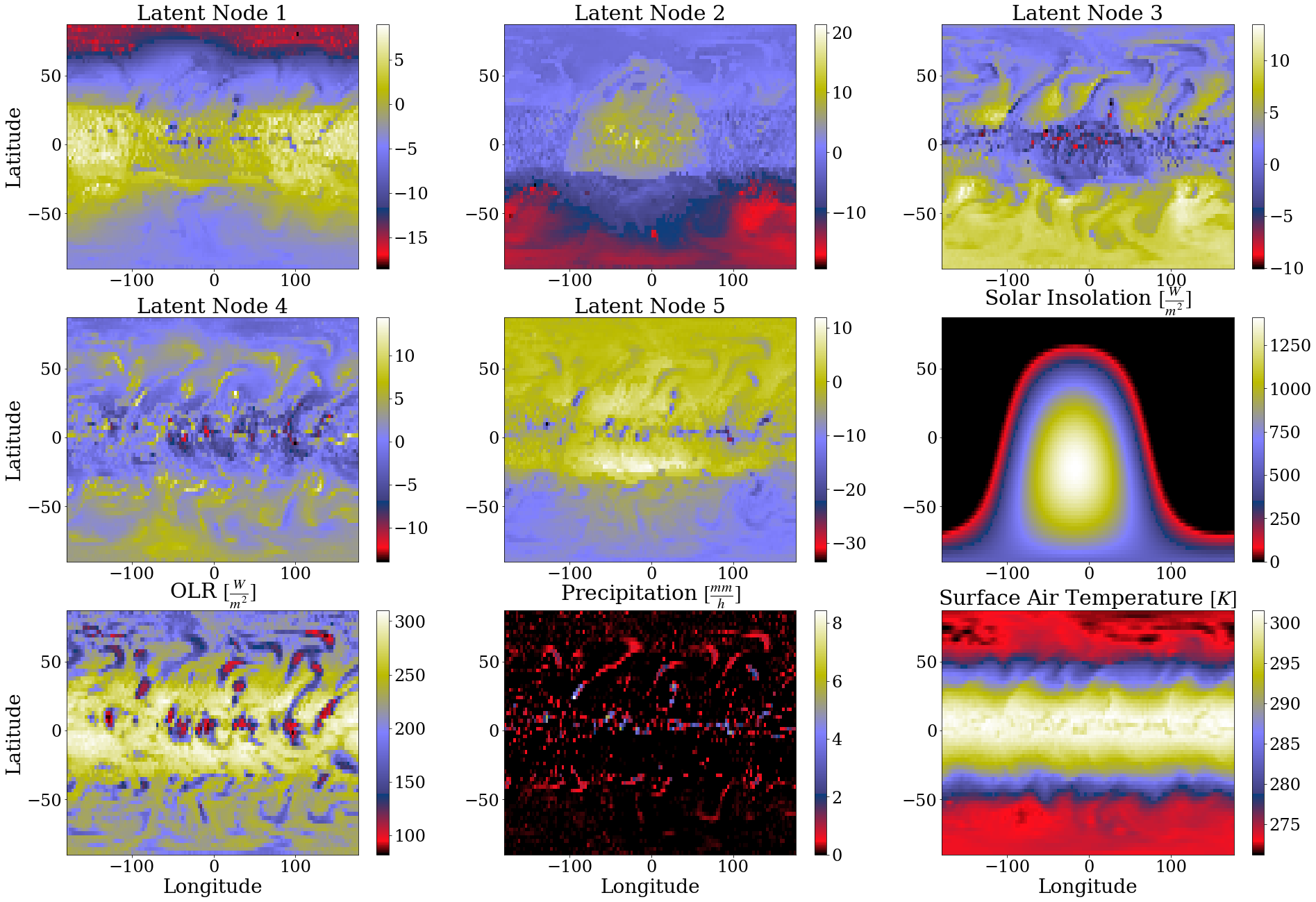}
    \caption{Latitude-Longitude plot of the latent variables of the VED (Latent Node 1 to 5) and respective large-scale and sub-grid-scale variables of the test data set for the same time step as in Figure \ref{fig:PCA_input_snap_shot}.}
    \label{fig:VAE_snaphot}
\end{figure}

\begin{figure}
    \centering
    \includegraphics[width=15.5cm]{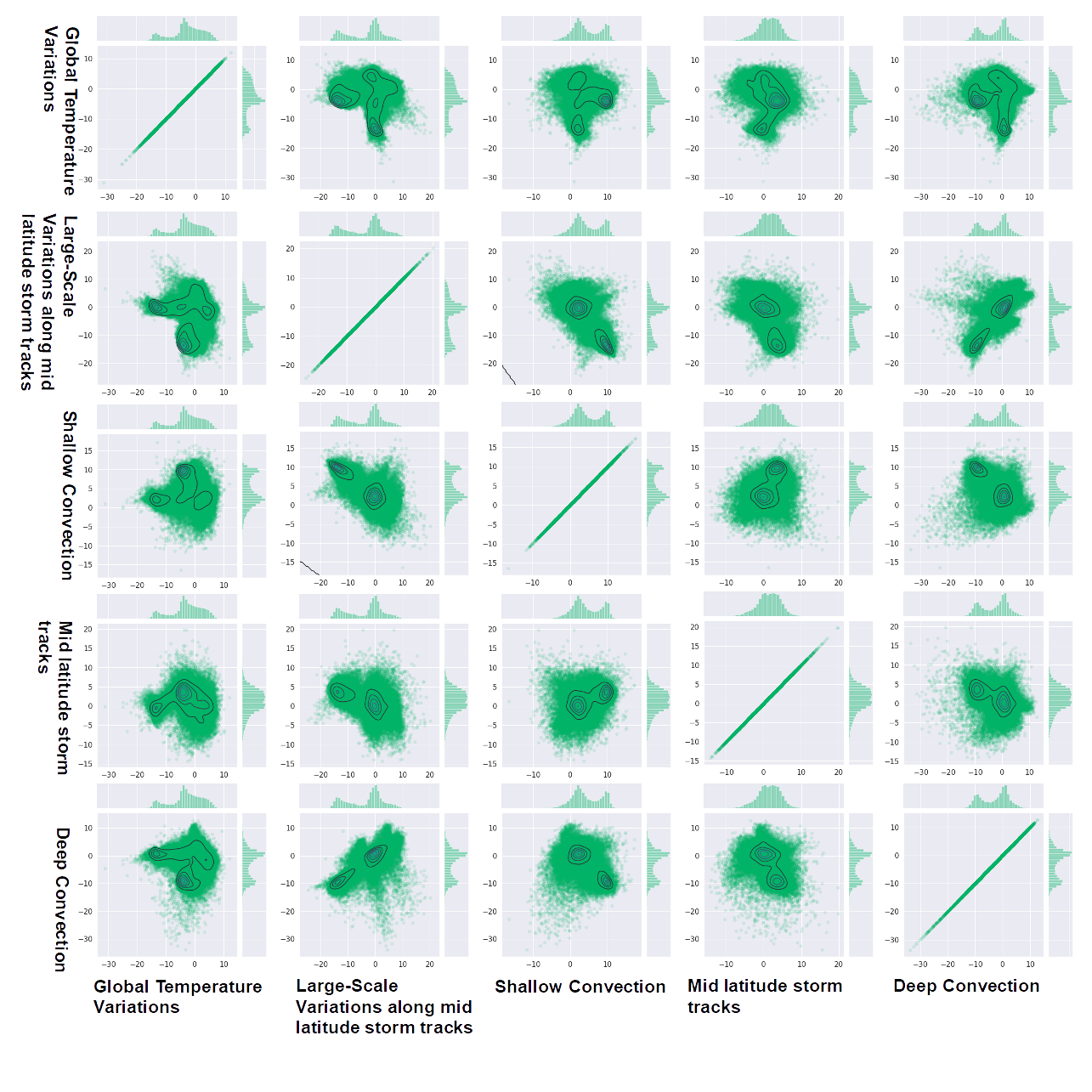}
    \caption{2D Density (blue contours) and scatter (light green) plots of the projection of two latent variables of the five dimensional latent space of VED in combination with the marginal distributions of the two latent variables. The first row represents the projection of Node 1 (Global temperature variations) onto all other four latent variables and itself. The second row shows the plots for Node 2 (Large-Scale variations along mid latitude storm tracks). The third row represents Node 3 (Shallow Convection). The forth and fifth row shows the plots for Node 4 (Mid latitude storm track) and Node 5 (Deep Convection). All plots are based on 50000 randomly selected samples from the test data.}
    \label{fig:latent_space_density_all_nodes}
\end{figure}

\begin{figure}
    \centering
    \includegraphics[width=15.8cm]{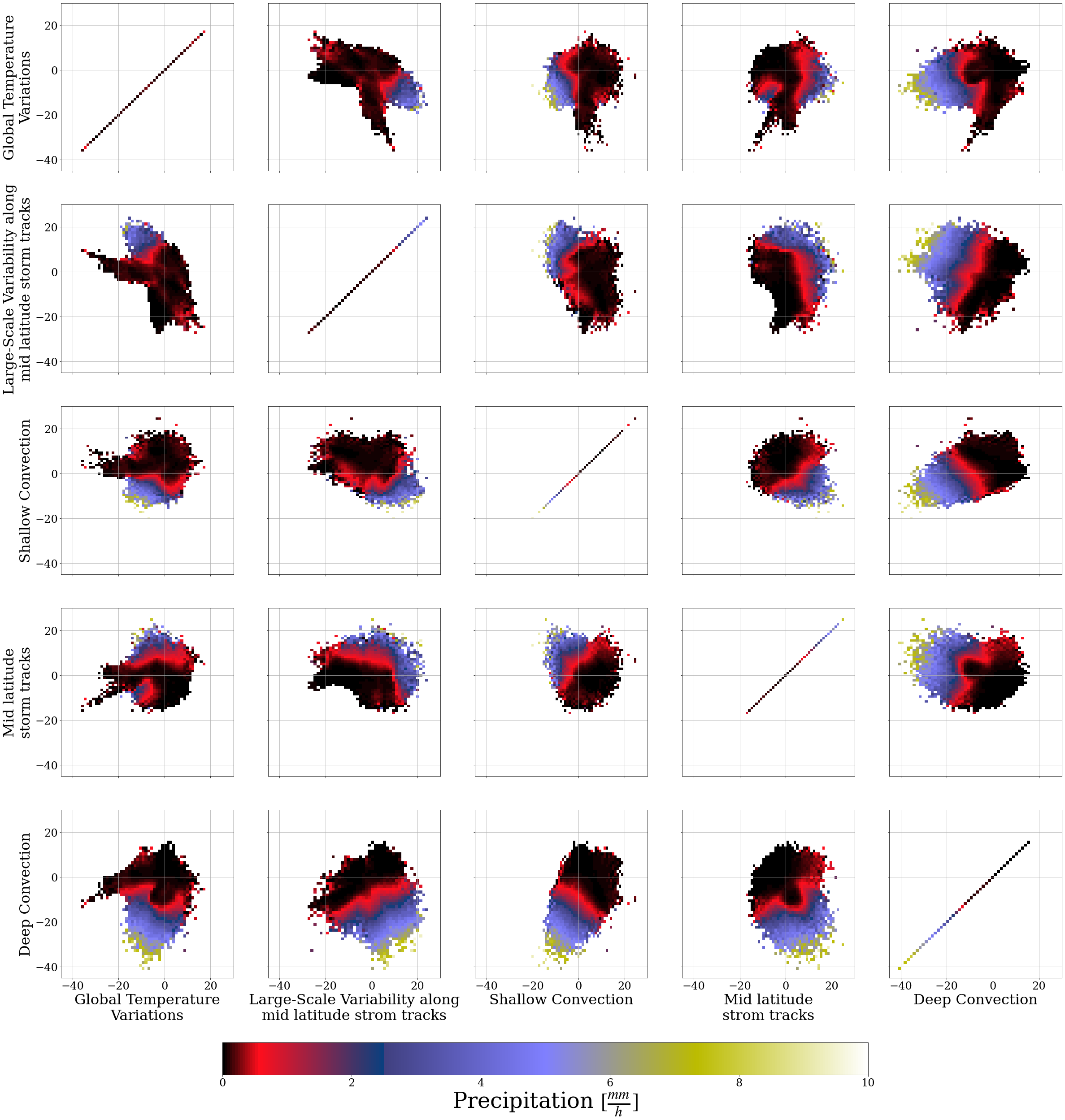}
    \caption{Similar to Figure \ref{fig:latent_space_density_all_nodes} a projection of one latent variable on all other latent variables and itself. The color coding reveals the conditional average of precipitation based on 1000000 randomly selected samples from the test data set. The first row shows the 2D projection of the first latent variable (Node 1, Global Temperature Variations) and all other latent variables. The second / third / fourth and fifth row depicts the projections of Node 2 (Large-Scale Variations along the mid latitude storm tracks) / Node 3 (Shallow Convection) / Node 4 (Mid latitude storm tracks) and Node 5 (Deep Convection).}
    \label{fig:latent_space_precip_cond_averages}
\end{figure}

\begin{figure}
    \centering
    \includegraphics[width=16.0cm]{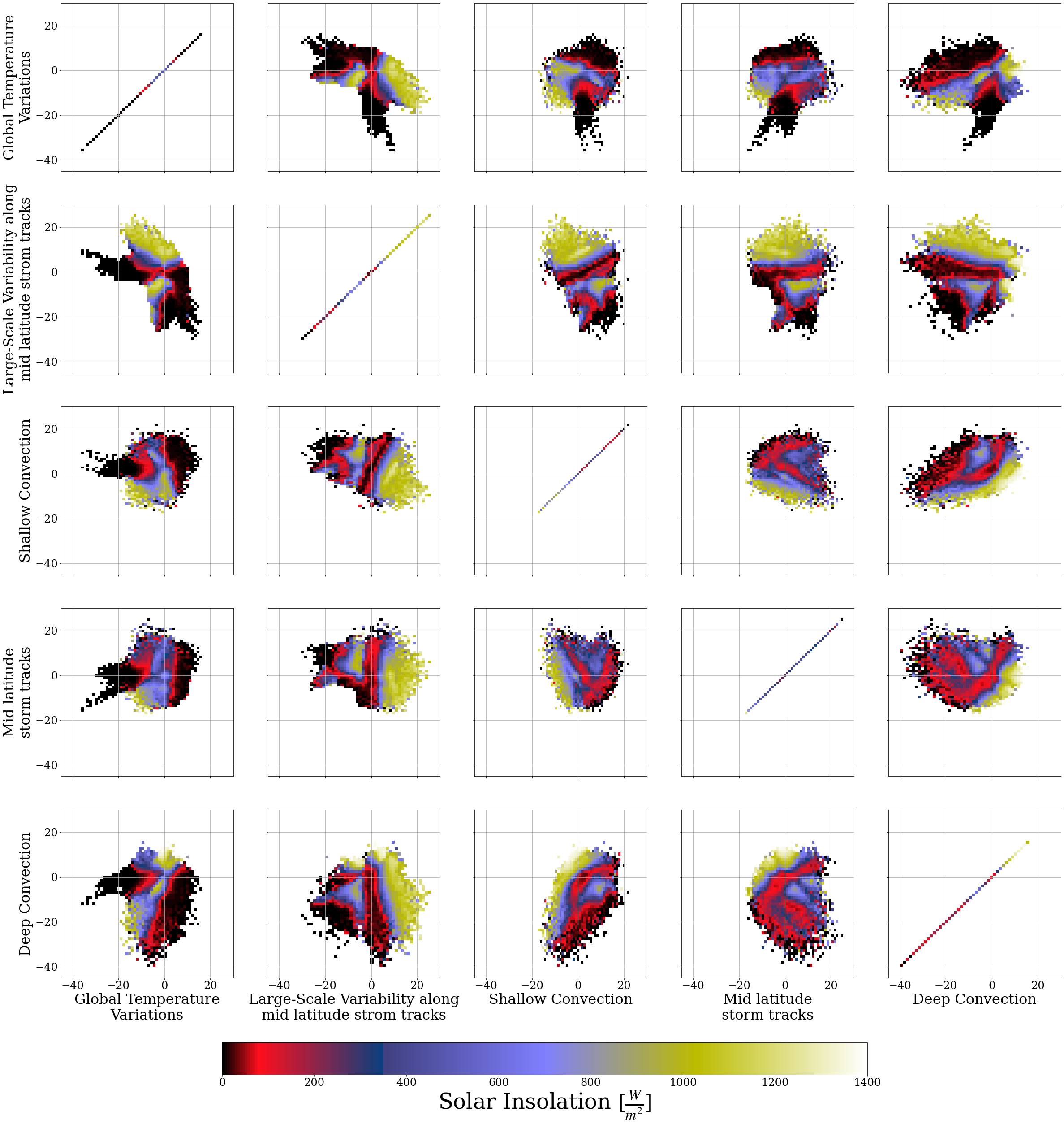}
    \caption{Similar to Figure \ref{fig:latent_space_precip_cond_averages}, but for the conditional averages of solar insolation in the projections.}
    \label{fig:latent_space_solar_insolation_cond_averages}
\end{figure}

\begin{figure}
    \centering
    \includegraphics[width=16.0cm]{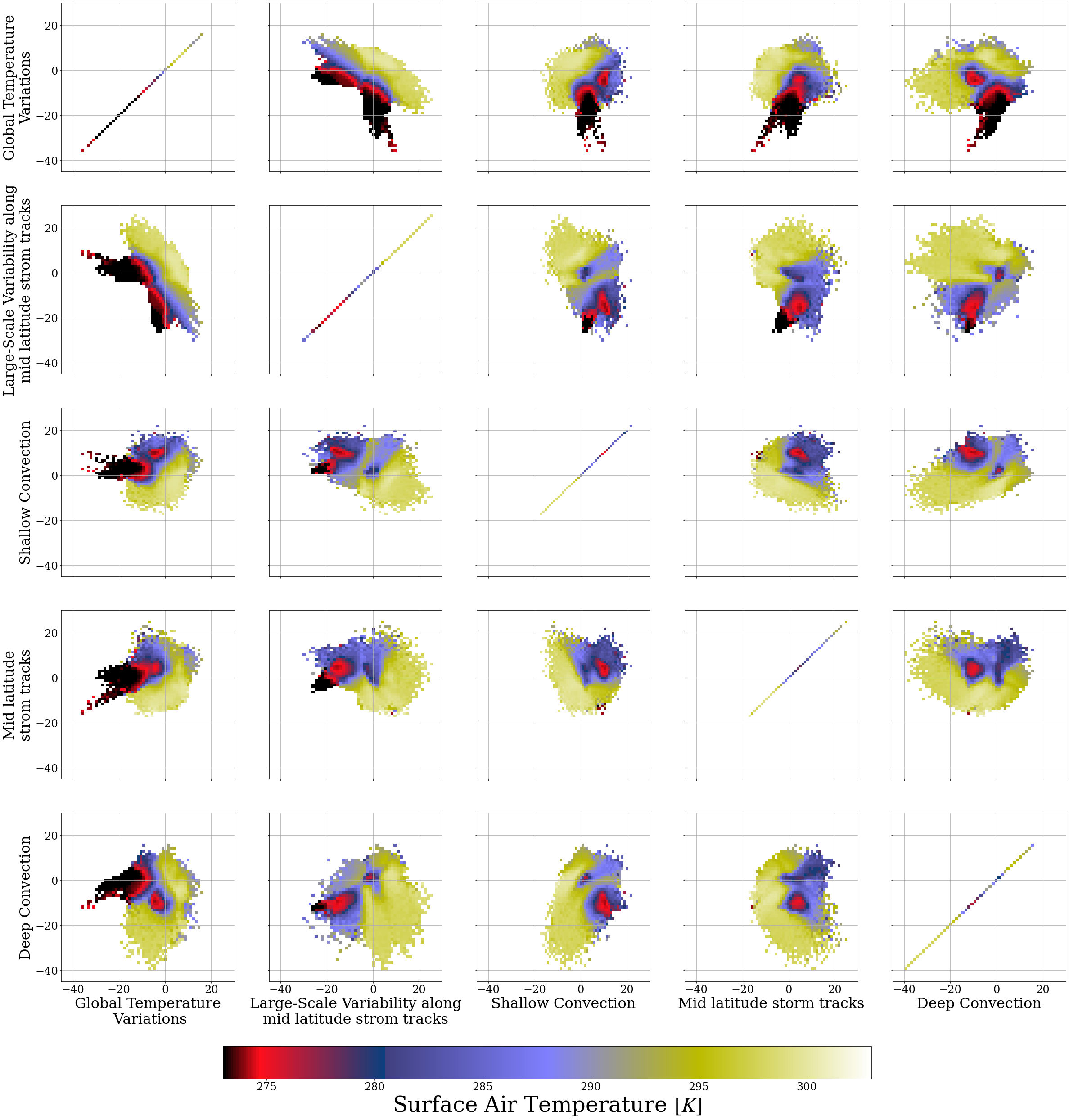}
    \caption{Similar to Figure \ref{fig:latent_space_precip_cond_averages}, but for the conditional averages of the surface air temperature in the projections.}
    \label{fig:latent_space_t_surf_cond_averages}
\end{figure}

The latent space of VED can be explored with the computation of conditional averages in a 2D PCA compressed submanifold as it is shown in Figure 5. However this analysis can be complemented with an inspection of the 5 latent dimensions itself. To visualize the five dimensional latent space, we projected two latent variables onto each other. This results in 20 spanned submanifolds of different latent variables and five projections of one latent variable onto itself oriented along the main diagonal in Figure \ref{fig:latent_space_density_all_nodes} to \ref{fig:latent_space_t_surf_cond_averages}. These 2D submanifolds of two different latent variables are often characterised by two or three centers of action with a strong concentration of samples (Figure \ref{fig:latent_space_density_all_nodes}). In most cases there is a weak linear connection between the latent variables, except for latent variable 2 (Large-scale variations along mid latitude storm tracks) and latent variable 5 (Deep Convection), as can be seen in Figure \ref{fig:latent_space_density_all_nodes}. The projection of these two latent variables is also characterised by a pronounced separation of samples with negligible convective processes (no precipitation, Figure \ref{fig:latent_space_precip_cond_averages}) and deep convective samples. This shows that the convective strength of the samples can be gauged with these two latent variables of VED and is not relying on a PCA as postprocessing step. Moreover the latent variables itself can be utilised to investigate large-scale geographic variability. One particular example for that, is the projection of latent variable 1 (Global temperature variations) and 2 (Large-scale variability along mid latitude storm tracks). In this submanifold we see two separated maxima of solar insolation (Figure \ref{fig:latent_space_solar_insolation_cond_averages}) and two areas with no solar insolation (night-time conditions). If we compare this distribution to the conditional averages of the surface air temperature (Figure \ref{fig:latent_space_t_surf_cond_averages}), we observe that one solar insolation maximum is associated with a minimum in surface air temperatures below 275K, which can be only observed in polar latitudes. The combination of solar insolation with anomalous cold temperatures is a clear evidence that the respective samples are originating from austral polar or subpolar latitudes due to the austral summer solar forcing of the SPCAM simulations. In contrast the other minimum in surface air temperatures in this projection of latent variable 1 and 2 is associated with no solar insolation. This suggests that the corresponding samples are coming from the boreal high latitudes (due to constant polar night conditions). These two examples illustrate that the interpretation of convective processes and large-scale drivers of convective predictability is possible on the latent variables of VED itself and not relying on the PCA postprocessing step. Furthermore the latent space of VED can be used to investigate longstanding hypotheses of atmospheric science. As an example we can focus on the projection of latent variable 5 (Deep Convection) onto latent variable 1 (Global temperature variations). The strongest precipitating samples in Figure \ref{fig:latent_space_precip_cond_averages} are situated in the middle of the conditional distribution of latent variable 1 (Figure \ref{fig:latent_space_density_all_nodes} and not in the right tail of the marginal distribution, which suggests that strong precipitation is not occurring in the regions with the highest surface air temperatures. This hypothesis can be evaluated with Figure \ref{fig:latent_space_t_surf_cond_averages}, where the region with the strongest precipitation is associated with conditional averages of surface air temperatures of around 295K in this projection. These temperatures are around 5K colder than the maximum of the conditional averages seen for this particular projection, which is in agreement with the original hypothesis. Overall these results indicate the power of the VED with respect to the interpretability and meaningfulness of the latent space and stored physical concepts in the lower-order manifold.

\noindent\textbf{S.3 Alternative VED and cVAE Structure}

\textbf{A)  VED$_{X \rightarrow Y}$}

\begin{figure}
    \centering
    \includegraphics[width=13.5cm]{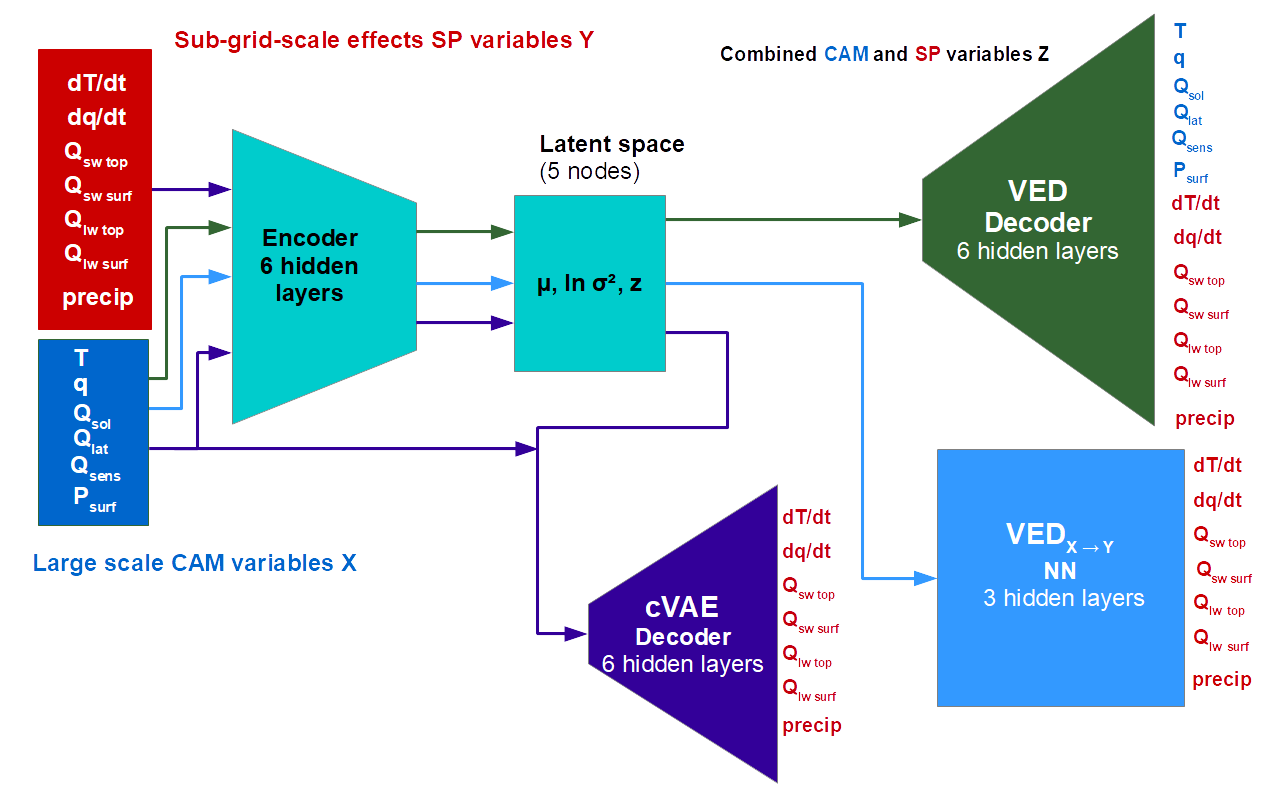}
    \caption{Combined schematic of the architecture of VED (green), VED$_{X \rightarrow Y}$ (light blue) and cVAE (purple arrows and network parts). The network structures in light blue are used for all variational networks with varying hyperparameters.}
    \label{fig:VAE_schematic}
\end{figure}

VED$_{X \rightarrow Y}$ closely mirrors the original SP with similar output variables to those of the reference ANN. It uses a set of convection related CAM climate variables \textbf{X} as input to the network, except for meridional wind profiles which were additionally used in \citeA{Rasp2018}. For this variational network, we couple the encoder to a regular feed forward neural net with 3 hidden layers. The resulting variational network VED$_{X \rightarrow Y}$ (Figure \ref{fig:VAE_schematic}) reproduces the convection-related SP output variables \textbf{Y} used in \citeA{Rasp2018}. The concatenated output vector \textbf{Y} has a length of 65 (65 output nodes). It contains the vertical profiles of temperature \textbf{dT(p)/dt} and specific humidity tendencies \textbf{dq(p)/dt}, the shortwave / longwave fluxes at the model top / surface \textbf{Q$_{sw/lw \ top/surf}$} and the precipitation rate \textbf{precip}. The coupled decoding feed forward neural net has three hidden layers with 353 nodes in each layer. The associated loss function is given in Equation 1.

\begin{equation}
    \mathrm{VED} \ \mathrm{loss_{X \rightarrow Y}}  =  \mathrm{reconstruction} \  \mathrm{loss_{X \rightarrow Y}} + {\lambda} \ \mathrm{KL} \ \mathrm{loss}
\end{equation}

 The reconstruction loss (Equation 2) of VED$_{X \rightarrow Y}$ is defined as the MSE between the emulated \textbf{Y$^{emul}$} and \textbf{Y}.
 
\begin{equation}
    \mathrm{reconstruction} \  \mathrm{loss_{X \rightarrow Y}} = {{ \frac{1} {M}} \times {\frac{1} {N}}} \sum_{i=1}^{M=65} \sum_{j=1}^{N= \mathrm{batch \, size}}(Y_{ij} - Y^{emul}_{ij})^2
\end{equation}

\begin{equation}
    \mathrm{KL} \  \mathrm{loss} = {{ \frac{1} {2}} \times {\frac{1} {N}}} \sum_{j=1}^{N=\mathrm{batch \,  size}} \sum_{k=1}^{K= \mathrm{latent \, space \, width}}\left[-1 - \ln \sigma^2_{jk} + \mu^2_{jk} +  \sigma^2_{jk}\right]
\end{equation}

\begin{equation}
    \lambda \ \epsilon \ \mathbb{R}_{+} 
\end{equation}

The hyperparameters used for VED$_{X \rightarrow Y}$ are displayed in Table \ref{tab:VAE_X_Y_hyper}, and the model architecture is illustrated in Figure \ref{fig:VAE_schematic}.

\begin{table}
    \centering
    \begin{tabular}{p{4cm}|p{8cm}}
         \textbf{Hyperparameter VED$_{X \rightarrow Y}$} & \textbf{Values}\\
         \hline
         \hline
         Learning Rate & 0.00018238 \\
         \hline
         Training / learning rate decrease & 40 epochs, learning decrease every 7$^{th}$ epoch by factor 5 \\
         \hline
         Batch size	& 714\\
         \hline
        Latent Space Width & 5 nodes \\
        \hline
        Node Size of Encoder &	[64,463,463,232,116,58,29,5]\\
        \hline
        Node Size of Decoder [ANN] &	[5,353,353,353,65]\\
        \hline 
        Activation Encoder &
        [Input, ReLU, ReLU, ReLU, ReLU, ReLU, ReLU, Lambda]\\
        \hline
        Activation Decoder [ANN] &
        [Input, ReLU, ReLU, ReLU, ELU]\\
        \hline 
        KL Annealing &	Linear annealing from 2$^{nd}$ to 7$^{th}$ epoch 
    \end{tabular}
    \caption{Hyperparameters and architecture of the constructed VED$_{X \rightarrow Y}$ which uses large scale CAM variables \textbf{X} to simulate SP variables \textbf{Y}.}
    \label{tab:VAE_X_Y_hyper}
\end{table}

\begin{figure}
    \centering
    \includegraphics[width=9cm]{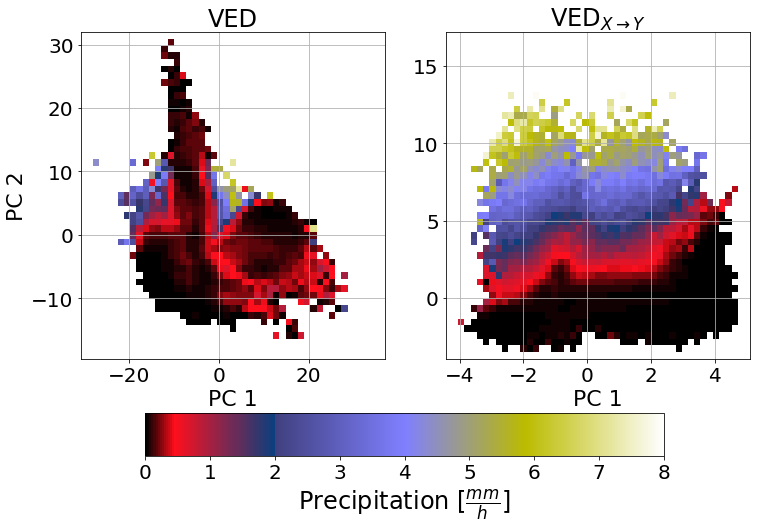}
    \caption{The 2D PCA compressed latent space of the VED (left) and VED$_{X \rightarrow Y}$ (right panel) and associated conditional average of precipitation of projected SP test data (similar to Figure 5). The x-axis / y-axis in all subplots indicates the 1$^{st}$/ 2$^{nd}$ leading PC of the 5D latent space in the respective panels.}
    \label{fig:VAE_VAE_X_Y_precip}
\end{figure}

\begin{figure}
    \centering
    \includegraphics[width=9cm]{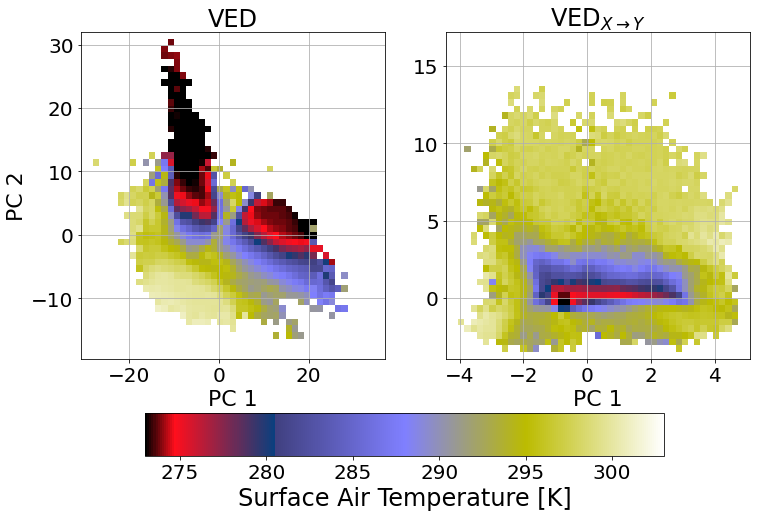}
    \caption{The 2D PCA compressed latent space of the VED (left) and VED$_{X \rightarrow Y}$ (right panel) and associated conditional average of surface air temperature of projected SP test data (similar to Figure 5). The x-axis / y-axis in all subplots indicates the 1$^{st}$/ 2$^{nd}$ leading PC of the 5D latent space in the respective panels.}
    \label{fig:VAE_VAE_X_Y_T_surf}
\end{figure}

VED$_{X \rightarrow Y}$ (test MSE = 0.157) reproduces the mean statistics with increased skill compared to VED (test MSE = 0.165) using the VED output normalization. The emulation skill of the spatio-temporal tropical variability is of the order of that of VED and slightly reduced with respect to reference ANN. However we see a decreased interpretability of the latent space of VED$_{X \rightarrow Y}$ in comparison to VED, which is a major disadvantage of the VED$_{X \rightarrow Y}$ network architecture. The 2D PCA compressed latent space of VED$_{X \rightarrow Y}$ generally shows a weak minimum to maximum distribution mostly focusing on the magnitude of convective processes (see Figure \ref{fig:VAE_VAE_X_Y_precip}) and faintly on geographic variability with respect to multiple sub-grid-scale and large-scale climate variables (see Figure \ref{fig:VAE_VAE_X_Y_T_surf}, as an example for surface air temperatures). Samples from the two poles with anomalously cold surface air temperatures are not well separated in the 2D PCA compressed latent space of VED$_{X \rightarrow Y}$, in contrast to that seen for VED (see Figure \ref{fig:VAE_VAE_X_Y_T_surf}). We observe one surface air temperature minimum in the 2D PCA compressed latent space of VED$_{X \rightarrow Y}$. The minimum comprises samples from the austral high latitudes to the right and from boreal latitudes to the left. These low surface air temperatures are compressed within a very small fraction of the 2D PCA compressed latent space of VED$_{X \rightarrow Y}$ surrounded by mid-latitude temperatures in close distance. For VED we see a clearly improved adaption to these large-scale meridional temperature variations with well separated zones of austral and boreal polar samples. Likewise we see for VED that samples with increased precipitation are concentrated into two centers of action and the 2D PCA compressed latent space illustrates strong gradients with respect to conditional averages of precipitation, which is not the case for VED$_{X \rightarrow Y}$ (Figure \ref{fig:VAE_VAE_X_Y_precip}). This lack of interpretability of the latent space is a general limitation of VED$_{X \rightarrow Y}$ compared to VED or even ED for the identification of driving large-scale climate conditions and related convective processes globally.

\textbf{B)  cVAE}

In general, a conditional VAE (cVAE) predicts the distribution of a set of output variables conditioned on the input variables. The general model configuration of cVAE's enables the propagation of information about the state of output variables and also input variables through the latent space to the conditional decoder \cite{Sohn2015}. For the task to realistically reproduce \textbf{Y} and gain insights on the interpretability of the latent space, we construct one possible cVAE. The sub-grid-scale variable vector \textbf{Y} is fed into the encoder together with large-scale CAM variables \textbf{X}, as can be seen in Figure \ref{fig:VAE_schematic}. \textbf{X} is an additional input to the decoding part of the network. As a result of that the latent space should illustrate a pronounced dependence on the sub-grid-scale input features \textbf{Y} rather than on large-scale CAM variables \textbf{X}. The cVAE's loss function is defined as:

\begin{equation}
    \mathrm{cVAE} \ \mathrm{loss}  =  \mathrm{reconstruction} \  \mathrm{loss_{cVAE}} + {\lambda} \ \mathrm{KL} \ \mathrm{loss}
\end{equation}

The associated reconstruction loss is defined as the MSE between \textbf{Y$^{emul}$} and \textbf{Y}, as can be seen in Equation 6. 

\begin{equation}
    \mathrm{reconstruction} \  \mathrm{loss_{cVAE}} = {{ \frac{1} {M}} \times {\frac{1} {N}}} \sum_{i=1}^{M=65} \sum_{j=1}^{N= \mathrm{batch \, size}}(Y_{ij} - Y^{emul}_{ij})^2
\end{equation}

\begin{equation}
    \mathrm{KL} \  \mathrm{loss} = {{ \frac{1} {2}} \times {\frac{1} {N}}} \sum_{j=1}^{N=\mathrm{batch \,  size}} \sum_{k=1}^{K= \mathrm{latent \, space \, width}}[-1 - \ln \sigma^2_{jk} + \mu^2_{jk} + \sigma^2_{jk}]
\end{equation}

\begin{equation}
    \lambda \ \epsilon \ \mathbb{R}_{+} 
\end{equation}

The used hyperparameters are displayed in Table \ref{tab:cVAE_hyper} and the model architecture can be seen in Figure \ref{fig:VAE_schematic}.

\begin{table}
    \centering
    \begin{tabular}{p{4cm}|p{8cm}}
         \textbf{Hyperparameter cVAE} & \textbf{Values}\\
         \hline
         \hline
         Learning Rate & 0.00096133 \\
         \hline
         Training / learning rate decrease & 40 epochs, learning decrease every 7$^{th}$ epoch by factor 5 \\
         \hline
         Batch size	& 666\\
         \hline
        Latent Space Width & 5 nodes \\
        \hline
        Node Size of Encoder &	[[65,64],457,457,228,114,57,29,5]\\
        \hline
        Node Size of Decoder  &	[5,29,57,114,228,457,457,65]\\
        \hline 
        Activation Encoder &
        [Input, ReLU, ReLU, ReLU, ReLU, ReLU, ReLU, Lambda]\\
        \hline
        Activation Decoder &
        [Input, ReLU, ReLU, ReLU, ReLU, ReLU, ReLU, ELU]\\
        \hline 
        KL Annealing &	Linear annealing from 2$^{nd}$ to 7$^{th}$ epoch 
    \end{tabular}
    \caption{Hyperparameters and architecture of the constructed cVAE which uses sub-grid-scale SP variables \textbf{Y} and large-scale CAM variables \textbf{X} to simulate sub-grid-scale SP variables \textbf{Y}.}
    \label{tab:cVAE_hyper}
\end{table}

Due to its deviating model architecture in comparison to the constructed VEDs (VED and VED$_{X \rightarrow Y}$), which are not trained with SP sub-grid-scale variables \textbf{Y} as input data, the cVAE has an advantage against all evaluated models in training mode (during the model optimization). This advantage in training mode reflects in a strongly improved emulation skill of this network compared to the reference ANN. The MSE of cVAE in training mode with respect to SP training, validation or test data (0.049 / 0.050 / 0.050) is more than half as small as the one of reference ANN (0.133 / 0.135 / 0.135) using the VED output normalization. We observe similar emulation capabilities for the related coefficients of determination R$^2$ of the lower tropospheric specific humidity and temperature tendencies. More than 96$\%$ of the horizontal grid points have a R$^2$ value larger than 0.7 for 700 hPa temperature tendencies in the case of the cVAE in training mode. For cVAE, only 38$\%$ of the grid points exceed a coefficient of determination of 0.7. Nevertheless the emulation capabilities in test mode, where only the CAM climate variables \textbf{X} are fed into cVAE, are remarkably weaker than for all other evaluated networks. This is one clear disadvantage of the ``brute-force training strategy'' of the cVAE with our architecture, where we train the encoder and decoder together. The strong decrease in emulation skill between training and test mode suggests that the largest portion of optimization goes into the emulation of the sub-grid-scale variables \textbf{Y}. Another discouraging point is the overall poorly developed interpretability of the latent space of cVAE with respect to essential sub-grid-scale and climate variables like outgoing longwave radiation, solar insolation or surface air temperature. cVAE is not capable to distinguish between day and night-time conditions in its latent space. This is a crucial benchmark of all other evaluated models. Overall, cVAE focuses in its latent space exclusively on variations in convective moistening and heating tendencies or the related formation of precipitation. This clearly limits the interpretability of drivers of convective predictability in the latent space of cVAE. Furthermore it suggests that key information about the background climate state of convective processes are dominantly propagated trough the additional link of \textbf{X} to the decoder of cVAE (see Figure \ref{fig:VAE_schematic}). This leads to the fact that the encoding of large-scale information in the latent space of cVAE in training mode is clearly outperformed by a traditional PCA on the climate variables \textbf{X}. Despite these discouraging results, we think cVAE could be upgraded towards a generative and stochastic parameterization of SP. \citeA{Pan2020} described that their initial cVAE structure exhibited large differences in performance between the training and test mode too. Therefore they developed a step-wise concept, where first the decoder is trained on \textbf{X}, then the encoder on \textbf{Y} and later the entire network on the complete variable list \textbf{O}. With this concept of step-wise training they were able to drastically improve the emulation abilities of their cVAE \cite{Pan2020}. In our case this upgraded training strategy might result in an enhanced interpretability of the latent space of cVAE with respect to large-scale drivers of convective predictability similar to results shown in this study for VED.

\noindent\textbf{S.4 Generated SP/CAM Variables with z$_{translation}$/z$_{median}$ and Squared Pearson Correlation R$^2$ Plots between Latent Nodes and Vertical Profiles} 

\begin{figure}
    \centering
    \includegraphics[width=16.0cm]{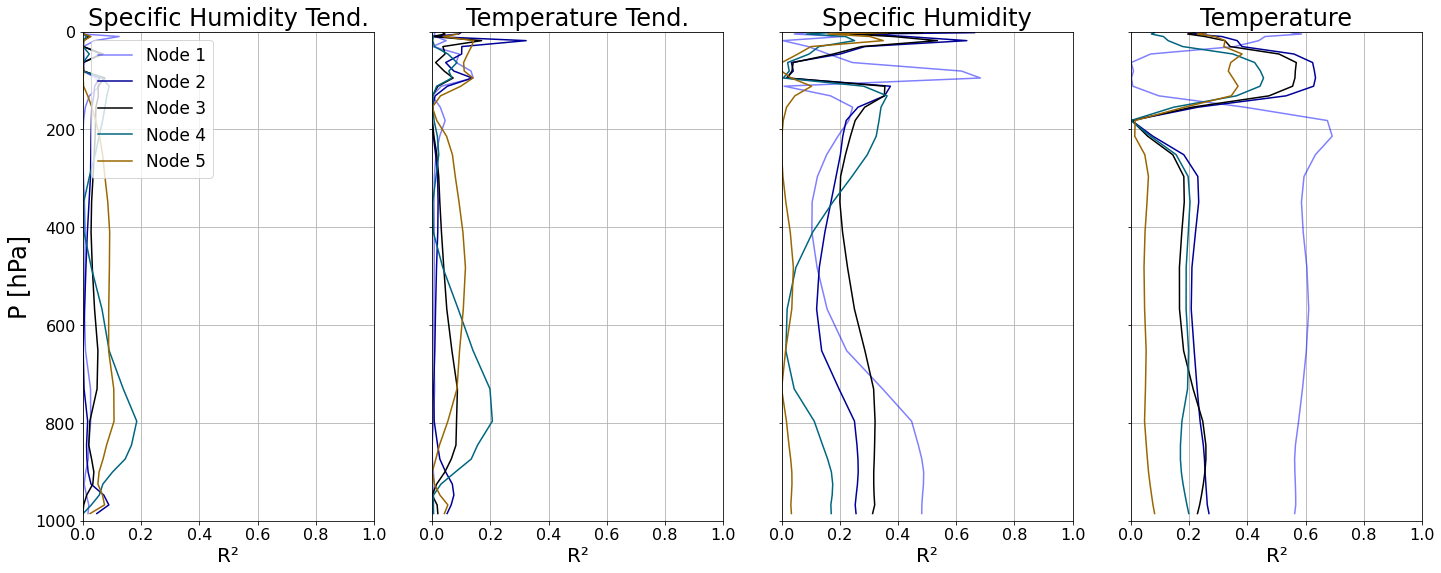}
    \caption{Squared Pearson correlation coefficient (linear explained variance) R$^2$ between the latent nodes of VED and predicted vertical profiles of specific humidity tendency (\textbf{dq/dt}), temperature tendency (\textbf{dT/dt}), specific humidity (\textbf{q}) and temperature (\textbf{T}) in space-time (which features the large meridional gradients of \textbf{q} and \textbf{T}). The light blue line resembles the R$^2$ value for latent node 1 / Global Temperature variations. The dark blue / black / dark cyan / bronze curve denotes the explained variance of latent node 2 (Large-scale variability along storm tracks) / 3 (Shallow Convection) / 4 (Mid latitude frontal system) / 5 (Deep Convection).}
    \label{fig:space_time_corr}
\end{figure}

\begin{figure}
    \centering
    \includegraphics[width=16cm]{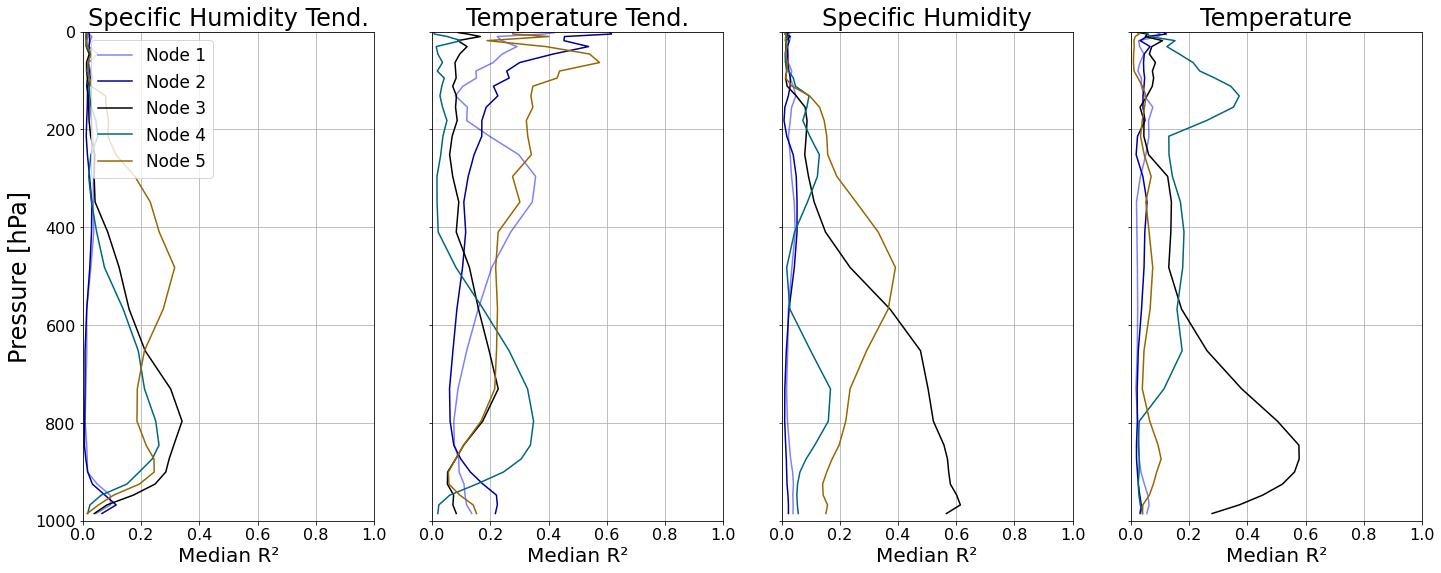}
    \caption{Median Squared Pearson correlation coefficient (linear explained variance) R$^2$ between the latent nodes of VED and predicted vertical profiles of specific humidity tendency (\textbf{dq/dt}), temperature tendency (\textbf{dT/dt}), specific humidity (\textbf{q}) and temperature (\textbf{T}) in time (without large meridional gradients of \textbf{q} and \textbf{T}). The light blue line resembles the median R$^2$ value for latent node 1 / Global Temperature variations. The dark blue / black / dark cyan / bronze curve denotes the median explained variance of latent node 2 (Large-scale variability along storm tracks) / 3 (Shallow Convection) / 4 (Mid latitude frontal systems) / 5 (Deep Convection).}
    \label{fig:time_corr}
\end{figure}

This section comprises the Tables \ref{tab:Node_1}-\ref{tab:Node_5} of generated 2D variables in \textbf{X} and \textbf{Y} for each latent node with our generative modeling approach. Additionally the squared Pearson correlation R$^2$ between the Nodes 1 to 5 and vertical profiles of \textbf{dq/dt}, \textbf{dT/dt}, \textbf{q} and \textbf{T} are displayed for space-time series (Figure \ref{fig:space_time_corr}) or time series (Figure \ref{fig:time_corr}) respectively. For Figure \ref{fig:space_time_corr} the Pearson correlation is computed based on the concatenated space-time  series (with the shape [horizontal grid-cells \textbf{H} $\times$ time steps \textbf{P}, latent space width \textbf{K} or output variable size \textbf{M}]) of the latent nodes and profiles in \textbf{O}, which means that these arrays include information about the large-scale geographic variability, e.g. the large meridional temperature and specific humidity contrasts between the tropics and poles. For Figure \ref{fig:time_corr} the Pearson correlation is calculated in each horizontal grid-cell between the time series (with the shape [\textbf{P}, \textbf{K} or \textbf{M}] of the latent nodes and output profiles in \textbf{O}. As a second step the median of the Pearson correlation coefficients is calculated across all horizontal grid-cells \textbf{H}.

\begin{table}
    \centering
    \begin{tabular}{p{4cm}||p{2cm}|p{2cm}|p{2cm}|p{2cm}|p{2cm}}
    \multicolumn{6}{c}{\textbf{Latent Node 1}}\\
    \hline
    \hline
    \textbf{Global Temperature variations} & 10$^{th}$ perc & 25$^{th}$ perc & 50$^{th}$ perc & 75$^{th}$ perc & 90$^{th}$ perc \\
    \hline
    \textbf{Q$_{sw \ top}$} [$\frac {W}{m^2}$] & 4 & 115 & 451 & 36 & 6\\
    \hline
    \textbf{Q$_{sw \ surf}$} [$\frac {W}{m^2}$] & -1 & 49 & 284 & 24 & 1 \\
    \hline
 	\textbf{Q$_{lw \ top}$} [$\frac {W}{m^2}$] & 181 & 221 & 241 & 260 & 275 \\
 	\hline
 	\textbf{Q$_{lw \ surf}$} [$\frac {W}{m^2}$] & 55 & 12 & 28 & 60 & 44 \\
 	\hline 
 	\textbf{precip} [$\frac {mm}{h}$] & 0.11 & 0.01 & 0.03 & 0.12 & 0.07 \\
 	\hline
 	\textbf{P$_{surf}$} [hPa] & 933 & 982 & 995 & 995 & 989 \\
 	\hline
 	\textbf{Q$_{sol}$} [$\frac {W}{m^2}$] & 15 & 263 & 748 & 45 & 8 \\
 	\hline
 	\textbf{Q$_{sens}$} [$\frac {W}{m^2}$] & 25 & 9 & 3 & 9 & 12 \\
 	\hline 
 	\textbf{Q$_{lat}$} [$\frac {W}{m^2}$] & 52 & 19 & 39 & 85 & 163
    \end{tabular}
    \caption{Generated shortwave and longwave heat flux at the model top / surface, precipitation, surface pressure, solar insolation, sensible and latent heat flux of \textbf{z$_{median}$} (4$^{th}$ column, 50$^{th}$ perc) and \textbf{z$_{translation}$} of the 10$^{th}$, 25$^{th}$, 75$^{th}$ and 90$^{th}$ percentile of latent node 1 (Global Temperature variations).}
    \label{tab:Node_1}
\end{table}

\begin{table}
    \centering
    \begin{tabular}{p{4cm}||p{2cm}|p{2cm}|p{2cm}|p{2cm}|p{2cm}}
    \multicolumn{6}{c}{\textbf{Latent Node 2}}\\
    \hline
    \hline
    \textbf{Large-scale variability along the mid latitude storm tracks} & 10$^{th}$ perc & 25$^{th}$ perc & 50$^{th}$ perc & 75$^{th}$ perc & 90$^{th}$ perc \\
    \hline
    \textbf{Q$_{sw \ top}$} [$\frac {W}{m^2}$] & 987 & 1092 & 451 & 57 & 158\\
    \hline
    
    \textbf{Q$_{sw \ surf}$} [$\frac {W}{m^2}$] & 773 & 845 & 284 &	30 & 49 \\
    \hline
 	\textbf{Q$_{lw \ top}$} [$\frac {W}{m^2}$] & 252 & 249 & 241 & 205 & 173 \\
 	\hline
 	\textbf{Q$_{lw \ surf}$} [$\frac {W}{m^2}$] & 85 & 73 & 28 & 44 & 13 \\
 	\hline 
 	\textbf{precip} [$\frac {mm}{h}$] & -0.01 & 0.00 & 0.03 & 0.12 & 0.15 \\
 	\hline
 	\textbf{P$_{surf}$} [hPa] & 983 & 989 & 995 & 993 & 992 \\
 	\hline
 	\textbf{Q$_{sol}$} [$\frac {W}{m^2}$] & 1214 & 1347 & 748 & 125 & 443 \\
 	\hline
 	\textbf{Q$_{sens}$} [$\frac {W}{m^2}$] & 19 & 9 & 3 & 6 & 23 \\
 	\hline 
 	\textbf{Q$_{lat}$} [$\frac {W}{m^2}$] & 83 & 55 & 39 & 86 & 101
    \end{tabular}
    \caption{Generated shortwave and longwave heat flux at the model top / surface, precipitation, surface pressure, solar insolation, sensible and latent heat flux of \textbf{z$_{median}$} (4$^{th}$ column, 50$^{th}$ perc) and \textbf{z$_{translation}$} of the 10$^{th}$, 25$^{th}$, 75$^{th}$ and 90$^{th}$ percentile of latent node 2 (Large-scale variability along mid latitude storm tracks).}
    \label{tab:Node_2}
\end{table}

\begin{table}
    \centering
    \begin{tabular}{p{4cm}||p{2cm}|p{2cm}|p{2cm}|p{2cm}|p{2cm}}
    \multicolumn{6}{c}{\textbf{Latent Node 3}}\\
    \hline
    \hline
    \textbf{Shallow Convection} & 10$^{th}$ perc & 25$^{th}$ perc & 50$^{th}$ perc & 75$^{th}$ perc & 90$^{th}$ perc \\
    \hline
    \textbf{Q$_{sw \ top}$} [$\frac {W}{m^2}$] & 5 & 134 & 451 & 1200 & 1112\\
    \hline
    \textbf{Q$_{sw \ surf}$} [$\frac {W}{m^2}$] & 2 & 49 &	284 & 925 & 838 \\
    \hline
 	\textbf{Q$_{lw \ top}$} [$\frac {W}{m^2}$] & 178 & 220 & 241 & 251 & 255 \\
 	\hline
 	\textbf{Q$_{lw \ surf}$} [$\frac {W}{m^2}$] & 3 & 8 & 28 & 	73 & 72 \\
 	\hline 
 	\textbf{precip} [$\frac {mm}{h}$] & 0.08 & 0.05 & 0.03 & 0.04 & 0.05 \\
 	\hline
 	\textbf{P$_{surf}$} [hPa] & 985 & 999 & 995 & 991 & 991 \\
 	\hline
 	\textbf{Q$_{sol}$} [$\frac {W}{m^2}$] & 15 & 329 & 748 & 1488 & 1468 \\
 	\hline
 	\textbf{Q$_{sens}$} [$\frac {W}{m^2}$] & -15 & -11 & 3 & 30 & 51 \\
 	\hline 
 	\textbf{Q$_{lat}$} [$\frac {W}{m^2}$] & -33 &	-13 & 39 & 207 & 242
    \end{tabular}
    \caption{Generated shortwave and longwave heat flux at the model top and surface, precipitation, surface pressure, solar insolation, sensible and latent heat flux of \textbf{z$_{median}$} (4$^{th}$ column, 50$^{th}$ perc) and \textbf{z$_{translation}$} of the 10$^{th}$, 25$^{th}$, 75$^{th}$ and 90$^{th}$ percentile of latent node 3 (Shallow Convection).}
    \label{tab:Node_3}
\end{table}

\begin{table}
    \centering
    \begin{tabular}{p{4cm}||p{2cm}|p{2cm}|p{2cm}|p{2cm}|p{2cm}}
    \multicolumn{6}{c}{\textbf{Latent Node 4}}\\
    \hline
    \hline
    \textbf{Mid latitude frontal systems} & 10$^{th}$ perc & 25$^{th}$ perc & 50$^{th}$ perc & 75$^{th}$ perc & 90$^{th}$ perc \\
    \hline
    \textbf{Q$_{sw \ top}$} [$\frac {W}{m^2}$] & 440 &  456 & 451 & 435 & 442\\
    \hline
    \textbf{Q$_{sw \ surf}$} [$\frac {W}{m^2}$] & 317 & 317 & 284 & 266 & 270 \\
    \hline
 	\textbf{Q$_{lw \ top}$} [$\frac {W}{m^2}$] & 175 & 202 & 241 & 224 & 215 \\
 	\hline
 	\textbf{Q$_{lw \ surf}$} [$\frac {W}{m^2}$] & 59 & 53 & 28 & 39 & 43 \\
 	\hline 
 	\textbf{precip} [$\frac {mm}{h}$] & 0.00 & 0.00 & 0.03 & 0.15 & 0.25 \\
 	\hline
 	\textbf{P$_{surf}$} [hPa] & 1001 & 1000 & 995 & 990 & 986 \\
 	\hline
 	\textbf{Q$_{sol}$} [$\frac {W}{m^2}$] & 625 &	678 & 748 & 741 & 746 \\
 	\hline
 	\textbf{Q$_{sens}$} [$\frac {W}{m^2}$] & -5 & -4 & 3 & 8 & 15 \\
 	\hline 
 	\textbf{Q$_{lat}$} [$\frac {W}{m^2}$] & 29 & 27 & 39 & 75 & 97
    \end{tabular}
    \caption{Generated shortwave and longwave heat flux at the model top / surface, precipitation, surface pressure, solar insolation, sensible and latent heat flux of \textbf{z$_{median}$} (4$^{th}$ column, 50$^{th}$ perc) and \textbf{z$_{translation}$} of the 10$^{th}$, 25$^{th}$, 75$^{th}$ and 90$^{th}$ percentile of latent node 4 (Mid latitude frontal systems).}
    \label{tab:Node_4}
\end{table}

\begin{table}
    \centering
    \begin{tabular}{p{4cm}||p{2cm}|p{2cm}|p{2cm}|p{2cm}|p{2cm}}
    \multicolumn{6}{c}{\textbf{Latent Node 5}}\\
    \hline
    \hline
    \textbf{Deep Convection} & 10$^{th}$ perc & 25$^{th}$ perc & 50$^{th}$ perc & 75$^{th}$ perc & 90$^{th}$ perc \\
    \hline
    \textbf{Q$_{sw \ top}$} [$\frac {W}{m^2}$] & 206 & 577 & 451 & 172 & 7\\
    \hline
    \textbf{Q$_{sw \ surf}$} [$\frac {W}{m^2}$] & 80 & 327 & 284 & 109 & 0 \\
    \hline
 	\textbf{Q$_{lw \ top}$} [$\frac {W}{m^2}$] & 188 & 208 & 241 & 254 & 266 \\
 	\hline
 	\textbf{Q$_{lw \ surf}$} [$\frac {W}{m^2}$] & 24 & 	26 & 28 & 93 & 113 \\
 	\hline 
 	\textbf{precip} [$\frac {mm}{h}$] & 0.60  & 0.24 & 0.03 & 0.01 & -0.01 \\
 	\hline
 	\textbf{P$_{surf}$} [hPa] & 989 & 989 & 995 & 999 & 998 \\
 	\hline
 	\textbf{Q$_{sol}$} [$\frac {W}{m^2}$] & 489 &	1036 & 748 & 264 & 	4 \\
 	\hline
 	\textbf{Q$_{sens}$} [$\frac {W}{m^2}$] & 4 & 2 & 3 & 3 & 6 \\
 	\hline 
 	\textbf{Q$_{lat}$} [$\frac {W}{m^2}$] & 57 & 51 & 39 & 64 & 80
    \end{tabular}
    \caption{Generated shortwave and longwave heat flux at the model top / surface, precipitation, surface pressure, solar insolation, sensible and latent heat flux of \textbf{z$_{median}$} (4$^{th}$ column, 50$^{th}$ perc) and \textbf{z$_{translation}$} of the 10$^{th}$, 25$^{th}$, 75$^{th}$ and 90$^{th}$ percentile of latent node 5 (Deep Convection).}
    \label{tab:Node_5}
\end{table}

%
%


%
%
%
%
%

 \bibliography{si_behrens_22.bbl}

%
%
%
%
%

%
%
\end{article}
\clearpage